\documentstyle[12pt,aaspp4]{article}
\begin{document}

\newcommand{\lap}{$L_{38}^{-1/3}$}
\newcommand{\ergs}{\rm \su  erg \su s^{-1}}
\newcommand{\etal}{ {\it et al.}}
\newcommand{\porb}{ P_{orb} }
\newcommand{\Po}{$P_{orb} \su$}
\newcommand{\pdot}{$ \dot{P}_{orb} \,$}
\newcommand{\pot}{$ \dot{P}_{orb} / P_{orb} \su $}
\newcommand{\mm}{$ \dot{m}$ }
\newcommand{\mdot}{$ |\dot{m}|_{rad}$ }
\newcommand{\myr}{ \su M_{\odot} \su \rm yr^{-1}}
\newcommand{\msol}{\, M_{\odot}}
\newcommand{\ppp}{ \dot{P}_{-20} }
\newcommand{\cms}{ \rm \, cm^{-2} \, s^{-1} }
\newcommand{\pdott}{ \left( \frac{ \dot{P}/\dot{P}_o}{P_{1.6}^{3}}\right)}
\newcommand{\be}{\begin{equation}}
\newcommand{\ee}{\end{equation}}
\newcommand{\nn}{\mbox{} \nonumber \\ \mbox{} }
\newcommand{\ba}{\begin{eqnarray}}
\newcommand{\ea}{\end{eqnarray}}

\def\p{\phantom{1}}
\def\pmu{\mox{$^{-1}$}}
\def\kms{km^s$^{-1}$}
\def\sbu{mag^arcsec${{-2}$}}
\def\e{\mbox{e}}
\def\dex{\mbox{dex}}
\def\L{\mbox{${\cal L}$}}
\def\gte{\lower 0.5ex\hbox{${}\buildrel>\over\sim{}$}}
\def\lte{\lower 0.5ex\hbox{${}\buildrel<\over\sim{}$}}
\def\loe{\lower 0.6ex\hbox{${}\stackrel{<}{\sim}{}$}}
\def\goe{\lower 0.6ex\hbox{${}\stackrel{>}{\sim}{}$}}

\title{
Electrodynamics of Magnetars: \break
Implications for the Persistent X-ray Emission and Spindown of \break
the Soft Gamma Repeaters and Anomalous X-ray Pulsars}

\author{
C. Thompson\altaffilmark{1},
M. Lyutikov\altaffilmark{2,3,4}, and
S.R. Kulkarni\altaffilmark{5}
}
\vskip 1in
\altaffiltext{1}{Canadian Institute for Theoretical Astrophysics,
60 St. George St., Toronto, ON M5S 3H8}
\altaffiltext{2}{Department of Physics, McGill University,
Montr\'eal, QC}
\altaffiltext{3}{Massachusetts Institute of Technology,
77 Massachusetts Avenue, Cambridge, MA 02139}
\altaffiltext{4} {CITA National Fellow}
\altaffiltext{5}{California Institute of Technology, 105-24, Pasadena,
CA 91125}

\vskip 1in
\centerline{Submitted to the Astrophysical Journal}

\begin{abstract}
We consider the structure of neutron star magnetospheres threaded
by large-scale electrical currents, and the effect of resonant
Compton scattering by the charge carriers (both electrons and ions)
on the emergent X-ray spectra and pulse profiles.  In the magnetar
model for the Soft Gamma Repeaters and Anomalous X-ray Pulsars,
these currents are maintained by magnetic stresses acting
deep inside the star, which generate both sudden disruptions
(SGR outbursts) and more gradual plastic deformations of the rigid
crust.  We construct self-similar force-free equilibria
of the current-carrying magnetosphere with a power law 
dependence of magnetic field on radius, ${\bf B} \propto r^{-(2+p)}$, 
and show that a large-scale twist of field lines
 softens the radial dependence of the magnetic field 
to $p < 1$.  The spindown torque acting on the
star is thereby increased in comparison with an orthogonal vacuum dipole.
We comment on the strength of the surface magnetic field
in the SGR and AXP sources, as inferred from their measured spindown rates,
and the implications of this model for the narrow measured distribution
of spin periods.

A magnetosphere with a strong twist ($B_\phi/B_\theta = O(1)$
at the equator) has an optical depth $\sim 1$ to resonant cyclotron
scattering, independent of frequency (radius), surface
magnetic field strength, or charge/mass ratio of the scattering charge.  
When electrons and ions supply the current, the stellar surface is
also heated by the impacting charges at a rate comparable to the observed
X-ray output of the SGR and AXP sources, if $B_{\rm dipole} \sim 10^{14}$ G.
Redistribution of the emerging X-ray flux at the cyclotron resonance
will strongly modify the emerging pulse profile and, through
the Doppler effect, generate a non-thermal tail to the X-ray spectrum.
We relate the sudden change in the pulse profile of SGR 1900$+$14 following
the 27 August 1998 giant flare, to an enhanced optical depth 
at the electron cyclotron resonance 
resulting from a sudden twist imparted to the external magnetic field 
during the flare.  The self-similar structure of the magnetosphere 
should generate frequency-independent profiles;
more complicated pulse profiles may reflect the presence of
higher multipoles, ion cyclotron scattering, or possibly non-resonant 
Compton scattering of O-mode photons by pair-loaded currents.

\end{abstract}

Subject Headings:
{gamma rays: bursts -- stars: neutron -- X-rays: stars}

\section{Introduction}

Our view of neutron stars has been strongly influenced by radio pulsars.
The demographics and physical properties of isolated neutron stars have 
largely been derived from extensive pulsar searches, and associations of 
pulsars with supernovae.  Over the last decade, thanks to new high 
energy missions, there has been a growing recognition of diversity in the 
neutron star population.  Considerable attention has been focused on 
the radio-quiet neutron stars (RQNS), which often have a bright 
presence in the X-ray sky (Frail 1998).  Although the known RQNS are certainly 
heterogeneous, two groups of objects seem to show sufficient regularity 
to classify them: the Anomalous X-ray pulsars (AXPs) and the Soft 
Gamma-ray Repeaters (SGRs).  Both appear to be young objects given 
their location within supernova remnants and star-forming regions.

The AXPs have spin periods in the range $P = 6 - 12$ s,
characteristic ages $P/\dot P = 3\times 10^3 - 4\times 10^5$ yr, and
X-ray luminosities $L_X = 5\times 10^{34} - 10^{36}$ erg s$^{-1}$ (for
recent reviews see Mereghetti 2000, Thompson 2001).  Three out of the
six AXPs lie close to the centers of SNRs, confirming their youth.  In
none of these objects is there evidence of a binary stellar companion,
or the usual indicators of accretion through a disk.  Their
spin-down luminosity is always less than a few percent of the observed X-ray 
luminosity.  The unfamiliar source of energy makes these objects 
anomalous.

SGRs are identified by their hard X-ray flares.  Once localized, their 
quiescent X-ray emission (also in the range $L_X \sim 5\times 
10^{34}-10^{36}$ erg/s) allows further studies of these equally enigmatic
objects.  SGRs have spin periods in the range 5-8 s but somewhat
smaller characteristic ages than the AXPs, less than a few
thousand years.   Only one or two of the four 
SGRs are concident with a SNR.  However, all the SGRs are found in 
star-forming regions, attesting to their youth.  It is important to note 
that only 50\% of pulsars with characteristic ages less than 
$\sim 5\times 10^4$ years have associated SNRs.  Thus, for both SGRs
and AXPs, the frequency of associations with SNRs appears no different
statistically from that of young pulsars (Gaensler et al.  2001).  
As with AXPs, there are no observational reasons to believe that these 
objects have companions; nor evidence for accretion.  The reader is 
referred to Hurley (2000) and Thompson (2000) for recent reviews of the SGRs.

The giant flare of 5 March 1979 (Mazets et al. 1979) now associated with 
SGR 0525$-$66 provided the first observational evidence 
that the SGRs are highly magnetized neutron stars or ``magnetars''.  These
objects where hypothesized to form through dynamo action in supernova collapse
(Duncan \& Thompson 1992).  Superstrong magnetic fields, $B > 10^{15}$G,
are able to supply enough energy to power the bursts, thereby 
matching the extreme peak luminosities of more a million times Eddington;
and to confine a significant proportion of the cooling plasma over the 
long ($> 200$ s) duration of the flare (Duncan \& Thompson 1992; Pacyz\'nski 
1992; Thompson \& Duncan 1995).   The presence of 8-s pulsations in the 
March 5 burst provided a consistent and independent support for the 
magnetar hypothesis.

The discovery of long period pulsations in the quiescent X-ray emission
of SGRs by Kouveliotou et al. (1998) has opened a rich field of timing
studies of SGRs (Woods et al. 1999c, 2001b; Kaspi et al. 1999, 2001;
Gavriil \& Kaspi 2001), 
from which we are now able to obtain critical information about both the
electrodynamics and the superfluid behavior of these interesting
objects, in much the same way as has been the case with timing studies
of radio pulsars.  

The same magnetar model which has successfully described the bright
outbursts of the SGRs is also able to account for their long
spin periods and their quiescent emission (through magnetic field decay).
Indeed, the SGRs in their quiescent states overlap with the AXPs
in a three-dimensional parameter space ($P$, $\dot{P}$ and $L_X$).
This striking similarity motivated the suggestion that they share a 
common energy source:  the decay of a very strong magnetic field 
(Thompson \& Duncan 1996, hereafter TD96).  The bursting activity of the SGRs
is highly intermittent, so that some of the AXPs may be dormant SGRs.  

The circumstantial connection between the SGRs and AXPs has become stronger
with two objects SGR 0526$-$66 and AXP 1E 1048.1$-$5937 constituting as
possible ``missing links'.  After the 5 March 1979 giant flare, SGR
0526-66 continued to emit a dozen shorter bursts, but has not been an
active burster since 1983.  Its persistent spectrum is consistent with
a pure power law with photon index \footnote{Here $dN/dE \propto
E^{-\Gamma}$.} $\Gamma = 3.5$ (Kulkarni et al. 2001).  This relatively
soft power-law index is characteristic of the AXPs, but is much softer
than the $\Gamma \simeq 2.2$ measured in the persistent emission of the
other SGRs.  By contrast, 1E 1048.1-5937 has never been observed to
burst, but its power-law index $\Gamma = 2.5$ is harder than that of
SGR 0526-66. Equally significantly, this object appears to have the
largest timing noise of AXPs and at levels similar to those of
SGRs (Kaspi et al. 2001).  Specifically, between 1994 and 1996, the
spindown torque of 1E 1048.1-5937 appears to have increased by a factor
$\sim 2$ over its long term value (Paul et al. 2000).  This behavior is
similar to the accelerating spindown observed in SGR 1900+14 several
months following the August 27 giant flare (Woods et al. 2001b).  Thus,
the latest timing results are unveiling a remarkable similarly in the 
detailed spin behavior of the SGRs and AXPs.

Within the combined population of SGR and AXP sources, there is overall
a positive correlation between spindown rate and the hardness of the
spectral index $\Gamma$ in the persistent emission (Marsden \& White
2001).
There is, similarly, a positive correlation between hardness and
pulsed fraction within the AXP population (e.g. Table 2 in Kaspi et al.
2001).

This smooth continuity of the X-ray and timing properties, when
combined with the other observational similarities summarized
above, provides a strong {\it phenomenological} link between SGRs and 
AXPs.  If the SGRs are magnetars, then so are the AXPs.

The discovery of optical counterparts of AXPs (Hulleman, van Kerkwijk 
\& Kulkarni 2000; Hulleman et al. 2001) offers new clues about
AXPs. The existing optical data on SGRs are consistent with SGR
counterparts being similarly faint (Kaplan et al. 2001a). These
authors find that the dimness of the optical counterpart is incompatible
with standard accretion disk models and thus by elimination provide
evidence for the magnetar model.

There do, however, exist significant differences between AXPs and SGRs.
Beyond the obvious difference -- the spectacular X-ray outbursts
of the SGRs -- there is also a puzzling discrepency between different
methods of estimating their relative ages.   The location of three AXPs 
close to the centers of their SNRs would suggest that -- given similar
proper motions -- the SGRs must descend from the AXPs. 
However, the characteristic ages of SGRs are typically smaller than
those of AXPs.  Combining these two observations, one infers
that the measured spindown of an SGR must be
transiently {\it accelerated} compared with its long-term average.
Precisely this effect is now being observed (over relatively brief
1-2 year intervals) in the timing solutions (Paul et al. 2000;  Woods 
et al. 2001b).  The positive correlation between spindown rate and 
cumulative bursting activity further suggests that activity as an SGR source
is also intermittent.  One of the basic goals of this paper is to connect 
this apparent transient behavior to the magnetar model.  

This general picture does not require a temporal link between the SGRs and
AXPs, but if one exists then the SGRs are probably old AXPs which
hibernate most of the time and display SGR-like activity
over a small fraction ($\sim 10-25\%$) of their lifetime.
Alternatively, some AXPs may never emit bright X-ray outbursts, and
may be distinguished from the SGRs in some hidden attribute, e.g., 
the multipolarity of the magnetic field.  

How do the AXPs and SGRs fit into the grander picture of neutron 
star behavior?  Recent radio surveys have discovered 
pulsars with polar magnetic fields approaching $10^{14}\,$G (Camilo 2000),
continuous with the lower range of fields deduced from AXP spindown.
However, these objects appear to be no different
from other radio pulsars and in particular are not X-ray bright
(Pivovaroff et al. 2000).  This bifurcation in X-ray behavior suggests 
that the magnetar phenomenon (e.g. rapid magnetic field decay) does not
turn on until $B$ exceeds $10^{14}\,$G.  In particular, AXPs and radio
pulsars may be distinguished by the strength of the {\it internal} 
magnetic field.

In this paper, we present a physical model which synthesizes the persistent 
emission and torque behavior of the SGRs and AXPs, relates that 
behavior to the mechanism driving the SGR flares, and provides
a strong suggestion that these objects were born with very rapid 
rotation. The starting hypothesis is that the magnetic field is globally
twisted inside the star -- up to a strength about 10 times the external
dipole -- and is strong enough to twist up the external field (at intervals). 
In the magnetar model, these currents provide the most promising mechanism
for generating the non-thermal persistent emission, through magnetospheric
Comptonization and surface heating (TD96).   Direct evidence for
this effect comes from the observed change in the pulse profile of
SGR 1900+14 following the 27 August 1998 giant flare (Thompson et al. 2000).  
We explain how these persistent currents flowing outside the star
will heat its surface, and modify the emerging X-ray spectrum and 
pulse profile.  The external twist will also increase the electrical
current flowing across the speed of light cylinder, and therefore the 
spindown torque acting on the star.  A positive correlation between
spectral hardness and spindown rate is a natural consequence of this
model.  

The presence of an ultrastrong magnetic field, $B \sim 4\times 10^{15}$ G, 
in the deep interior of the SGR and AXP sources has a few further
interesting consequences.  

First, such a strong field can comfortably account for the total
energy released by these sources in various channels.
Not only do magnetars lose energy through electromagnetic emissions
(X-ray, UV and optical), but also through relativistic particles
and neutrino radiation from the deep crust and core.  There is evidence 
for particle losses during the 27 August 1998 giant burst of 
SGR 1900+14 (Frail, Kulkarni \&\ Bloom 1999).  While current observations
do not show any plerions around other SGRs (or AXPs), the upper
limits on the persistent particle luminosity obtained from these 
observations are not constraining.  If the observed X-ray output of
the AXPs is dominated by transport of the magnetic field through
the neutron star core, then neutrino losses will exceed the surface
flux by a factor $\sim 30$ (TD96; Heyl \& Kulkarni 1998).
Estimating the bolometric losses is also complicated in some sources
(primarily the AXPs) by a steep power-law component in the X-ray spectrum,
which is not observed below $\sim 0.5$ keV due to absorption.
Assuming a typical {\it observed} luminosity of $3\times 10^{35}\,$erg 
s$^{-1}$ and a mean age of $10^4\,$yr we obtain $E \sim 10^{47}\,$erg,
comparable to the inferred output of the bursting SGR sources.
This should be compared to the total magnetic energy of
$10^{47}B_{15}^2\,$ erg  where $B_{15}$ is the strength of the
field in units of $10^{15}\,$G and it is assumed that this
field permeates the entire volume of the neutron star.  The
very high electrical conductivities of neutron star interiors will
prevent currents from being entirely dissipated, and so internal
fields somewhat stronger than $\sim 10^{15}$ G are needed to supply
the entire bolometric output of an AXP or SGR source.
This simple estimate also underlines the importance of obtaining
more precise measurements of the low-energy spectral cutoff in 
sources with soft spectra.

Second, the presence of ultrastrong internal magnetic fields leads
to a simple physical mechanism for the termination of
the bright X-ray emissions of the SGR and AXP sources.  It is clear
observationally that the AXP/SGR phenomenon is restricted to young
objects, with total ages less than $10^4-10^5\,$yr.  What causes this
decline in magnetar activity?  One explanation involves the strong
temperature-dependence of the rate of magnetic field transport 
through the neutron star core (e.g. Goldreich \& Reisenegger 1992).  
Transport is dramatically accelerated by internal heating, and so the
transport of a deeply anchored field will cut off sharply when 
cooling of the star is dominated by surface X-ray emission (TD96).  

What do old magnetars look like?
It is in this context that the discovery of long period isolated
nearby X-ray pulsars such as RX~J0720.4$-$3125 (Haberl et al. 1997)
becomes interesting. There are three such objects (see Treves et al.
2000 for a review), a high fraction (30\%) of nearby neutron stars.
Several authors have argued these objects to be old magnetars
(e.g. Heyl \&\ Hernquist 1998, Kulkarni \&\ van Kerkwijk 1998).
If indeed this is the case, then the birthrate
of magnetars is a significant fraction of the pulsar birth rate.
Furthermore, the similarity of the periods of these objects with those 
of AXPs+SGRs indicates that period evolution may slow down after the active
magnetar phase.  

Third, as we have already mentioned, the presence or absence of a strong
toroidal field offers an explanation for the bifurcation in behavior between
radio pulsars and the X-ray bright SGRs and AXPs.  More generally, the 
maximum {\it total} (as opposed to dipolar) field which will manifest 
pulsar-like activity (and the minimum field which is needed to power an 
AXP or SGR) needs a more detailed theoretical explanation.  These minima
and maxima need not coincide.  Is there a gap between the B-field strengths
of the observed high-field pulsars, and the magnetar population?  The
theoretical model advanced in this paper will be useful in making this 
question more precise.

\subsection{Plan of the Paper}

In this paper, we consider the properties of neutron stars whose
external magnetic fields support {\it global} electrical currents,
flowing across a large fraction of the stellar surface.
We construct axisymmetric, self-similar
solutions to the force-free equation, which describe the structure of
a magnetosphere with a global twist between the two hemispheres.
We begin in Section 2 with a general discussion of how the decay of
an electrical current flowing inside stellar material of a very high
electrical conductivity may be accelerated as hydromagnetic stresses
divert part of the current to the exterior, where it is damped much more
rapidly.  Section 3 describes the construction and basic properties of
our self-similar model, which forms a one-parameter sequence labeled
by the net angle through which the magnetic field lines are twisted
(or, equivalently, by the radial power-law index of the magnetic field).
The modification of the corotation charge density by static currents
is calculated, and the matching of our twisted solution onto open
magnetic field lines is discussed qualitatively.

Resonant cyclotron scattering of photons by the charge carriers
is explored in Section 4, where the following remarkable property is
demonstrated:  the optical depth to resonant scattering is proportional
to the twist, but does not depend explicity on the radius, or the mass
and electrical charge of the current-carrying particles.  This resonant
optical depth is largest at the magnetic equator, but decreases to
zero near the magnetic axis (where the current flowing along more
extended field lines is relatively weak).

Section 5 gives an introduction to how the X-ray spectrum and pulse
profile of a magnetar will be modified by multiple resonant scattering
in the magnetosphere, as well as the implication for the mechanism of
the giant flares in the SGR sources.
In Section 6, we calculate how the surface field inferred from spindown
measurements is modified, and then consider the implications of our
model for the narrow distribution of pulse periods measured in the SGR
and AXP sources.  Section 7 closes the paper with a summary of our
findings,
and their more general implications for the nature of the SGR and AXP
sources.

\section{Twisted Neutron Star Magnetospheres}

The non-thermal persistent emission of the SGRs has been ascribed
to a static twist imparted to the external magnetic field by 
sub-surface motions during X-ray flares, with the effect of
diverting an electrical current from the interior of the star to its
exterior (Thompson et al. 2000).   Magnetic fields exceeding
$\sim 10^{14}$ G are strong enough to fracture the deep crust of a neutron
star and, if stronger than $\sim 10^{15}$ G, will undergo rapid transport 
through the dense stellar interior over the short 
$\sim 10^4-10^5$ yr active lifetime of the SGR/AXP sources 
(TD96; Heyl \& Kulkarni 1998).  

The magnetic fields of neutron stars are most likely generated by
a hydromagnetic dynamo as the star is born.
At this time, magnetic fields are easily transported across the
boundary of the neutron-rich core, where they can reconnect
from the field anchored in the star (Thompson \& Murray 2001).  In this
manner, net magnetic helicity builds up in the anchored field
as the disconnected field carries away helicity.   Note also that the
net twist which is maintained along extended field lines
(reaching out beyond several stellar radii) is controlled
by the current flowing through only a fraction of the surface.

\begin{figure}
\plotone{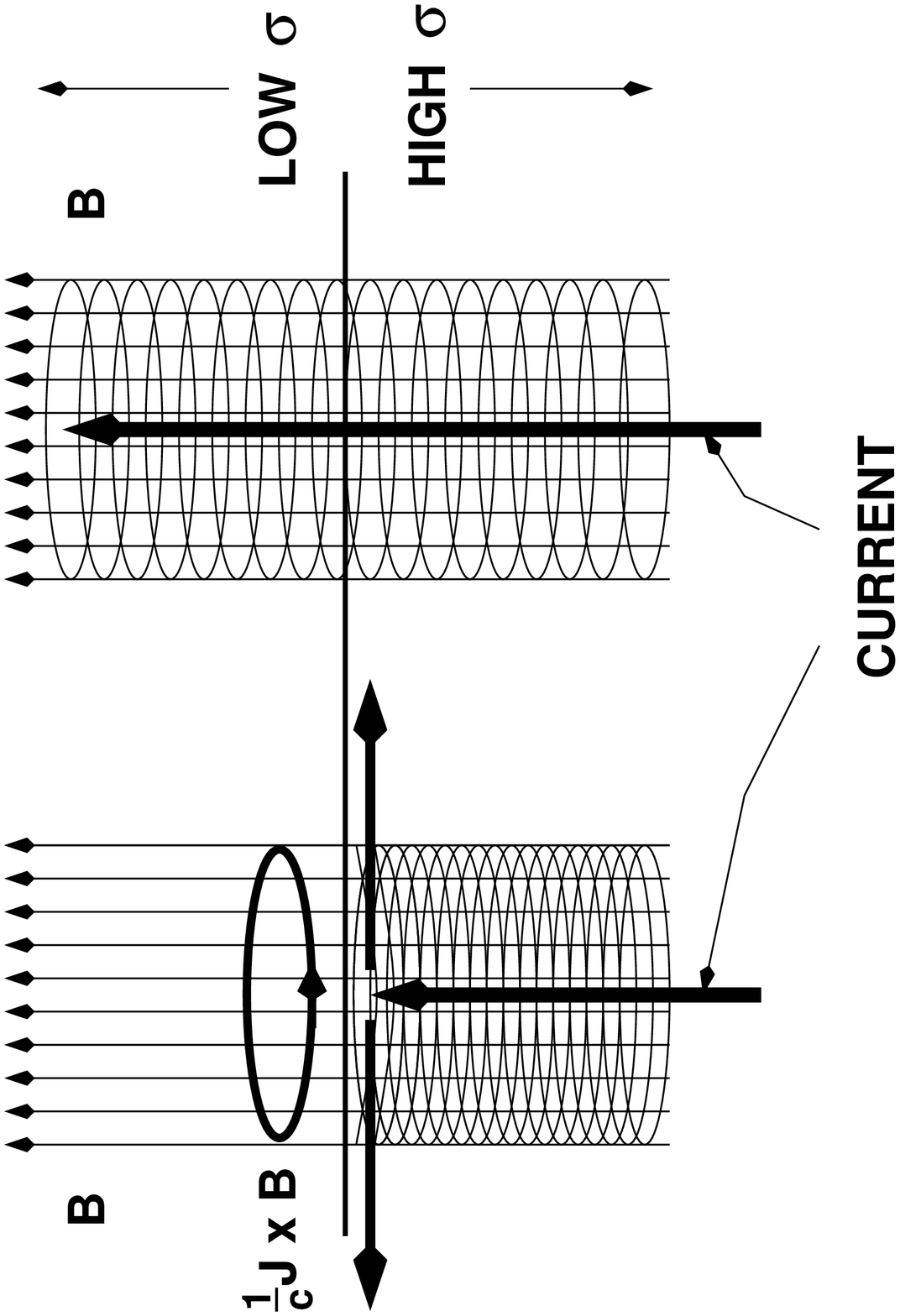}
\label{liquid}
\end{figure}

Consider a twisted magnetic field which is anchored in a highly conducting 
crust of a neutron star (Fig. \ref{liquid}).    Suppose that initially 
the twist vanishes outside the  star, where the conductivity is 
much lower.  Then the current closes through a thin surface layer,
which feels a Lorentz force ${1\over c}{\bf J}\times {\bf B}$.  In the
absence of any tensile strength, this force causes the surface layer and
the external field to twist up.  The net effect is to distribute the 
twist more uniformly along the flux tube (Fig. \ref{liquid}), and to 
hasten the decay of the current by forcing it into a medium of lower 
conductivity -- the magnetosphere.  The twist will then relax in a 
time comparable to the decay time $t_{\rm decay}^{\rm ex}$ {\it outside} 
the star.  In the case of a realistic neutron star, this process will be 
impeded by the compositional stratification of the stellar core:
the untwisting of the magnetic field is suppressed if it requires the 
motion of fluid elements across equipotential surfaces (Thompson 
\& Duncan 2001).

In the SGRs and AXPs, such an external
current can be supported by charges stripped from the surface:   the
effective temperature is high enough to allow thermal emission of electrons 
and light ions (hydrogen and helium), even in the presence of a 
$\sim 10^{15}$ G magnetic field (Thompson et al. 2000).  Any large-scale
deformation of a neutron star is further constrained by 
the rigidity of its crust.  A weak magnetic field is pinned by 
the crust, and the external current decays without being replenished.
The behavior of magnetic fields stronger than $\sim 10^{14}-10^{15}$ G
depends on the intermittency with which the crust responds to the 
applied magnetic stresses.  The compositional
stratification of a neutron star (Reisenegger \& Goldreich 1992)
strictly limits the degree to which
an internal magnetic field can unwind promptly.
The lowest energy deformations of the rigid crust are rotational,
and large-scale fractures can be triggered repeatedly as transport 
processes (such as ambipolar diffusion and Hall drift) allow the 
shear components of the magnetic stress to build up with time.
An empirical estimate of this growth time is given by the mean duration 
$\Delta t_{\rm flare}\sim 10^2$ yr between giant flares in any one
SGR source (e.g. Mazets et al. 1999).  
The external current will be maintained more or less
continuously if $t^{\rm ex}_{\rm decay} \ga t_{\rm growth} \sim 
\Delta t_{\rm flare}$;  
otherwise, the external current will decay in between spasmodic
events.  More gradual, plastic deformations of the star are also possible and,
in the magnetar model, are required to explain measured variations in the
X-ray output of some AXPs (e.g. Iwasawa, Koyama, \& Halpern 1992) which 
are not associated with bright X-ray bursts.  In this case, 
a non-potential magnetic field will be maintained outside the star 
only if $t^{\rm ex}_{\rm decay} \ga t_{\rm growth}$.

\section{Force-Free Equilibria}\label{force-free}

The exterior of an isolated neutron star is traditionally modeled
as a potential magnetic field, excluding a narrow bundle of
field lines which extend out to the speed of light
cylinder (Pacini 1967; Goldreich \& Julian 1969; Ruderman \& Sutherland 1975).
We generalize this classical dipole model to include an electrical 
current flowing continuously across the entire surface of the star.
We assume initially that the star does not rotate, so that the closed
magnetic field lines fill an infinite volume outside the star.  The 
effects of slow rotation are discussed in Sections \ref{roteffects} 
and \ref{Matching}.

The energy density of the charge carriers needed to supply the current
is, in this context, tiny compared with $B^2/8\pi$, and the bare space charge
$|\rho| \ll J/c$ on the closed field lines.  This leads us to solve
the force-free equation, ${\bf J}\times{\bf B} = 0$, to determine 
the structure of the magnetic field around a highly conducting 
spherical mass of radius $R_{\rm NS}$.  The solution of this equation 
can be written formally as
\begin{equation}\label{ff}
{\bf\nabla}\times{\bf B} = \alpha({\cal P}) {\bf B}.
\end{equation}
When the magnetosphere is axisymmetric (as we will assume), the
magnetic field lines form a one-dimensional sequence labeled by 
the flux parameter ${\cal P} = {\cal P}(R,\theta)$.  The poloidal
magnetic field can be written as 
\begin{equation}\label{bpval}
{\bf B}_P = {{\bf\nabla}{\cal P} \times {\hat\phi}\over R\sin\theta}.
\end{equation}
(In components, $B_R = -R^{-2}\partial{\cal P}/\partial(\cos\theta)$ and
$B_\theta = -(R\sin\theta)^{-1}\,\partial{\cal P}/\partial R$.)
The constancy of the current along an infinitesimal bundle of
magnetic flux guarantees that the coefficient $\alpha$ is a function
only of ${\cal P}$.
  As a major simplification we search for self-similar configurations 
\begin{equation}\label{stream}
{\cal P} = {\cal P}_0 r^{-p}F(\cos\theta),
\end{equation}
adapting the trick of Lynden-Bell and Boily (1994) who studied
force-free equilibria bounded by an infinite, conducting plane.
Here, $r = R/R_{\rm NS}$, $\theta$ is the magnetic co-latitude,
and ${\cal P}_0 = {1\over 2}B_{\rm pole}R_{\rm NS}^2$.  
Comparing with equation (\ref{ff}) shows that $\alpha({\cal P})$
is proportional to ${\cal P}^{1/p}$, and on dimensional grounds
one can write
\begin{equation}\label{alphaval}
\alpha({\cal P}) = {C^{1/2}\over R_{\rm NS}}\left({p+1\over p}\right)^{1/2}\,
\left({{\cal P}\over{\cal P}_0}\right)^{1/p}.
\end{equation}
The shape of the field lines, including the radial index $p$, is
determined uniquely by the single parameter $C$ which is related to the
strength of the current.   The poloidal components of
eq. (\ref{ff}) can be integrated to give
\begin{equation}\label{bphival}
B_\phi = {\int \alpha({\cal P})\,d{\cal P}\over R\sin\theta}
= {p\over p+1}\,{{\cal P}\alpha({\cal P})\over R\sin\theta}.
\end{equation}
Substituting eqs. (\ref{bpval})-(\ref{bphival}) into the $\phi$-component
of eq. (\ref{ff}) then gives the non-linear equation
\begin{equation}\label{gseq}
p(p+1)F + (1-\mu^2)
{\partial^2 F\over\partial\mu^2} = -CF^{1+2/p}
\end{equation}
for the angular factor $F = F(\mu)$.  

The solution of eq. (\ref{gseq}), including the dependence $p(C)$,
is uniquely defined by the parameter $C$ and by the three boundary
conditions: $B_R \propto F' = 0$ at 
$\mu \equiv \cos\theta = 0$ (the magnetic equator),  $F' = {\rm const}
= -2$ at $\mu = 1$ (corresponding to a fixed flux density $B_{\rm pole}$ 
at the magnetic pole), and $ B_{\phi}(\mu=1) \propto F(1)=0$.  
The solutions are symmetric under $\mu \leftrightarrow -\mu$,
with $F(\mu) = 1-\mu^2$ representing a pure dipole.  
For each value of $C\leq 0.873$ there
are two solutions for $p$.  The upper branch connects continuously to the 
vacuum dipole $C = 0$, $p = 1$ ($B_R, B_\theta \propto R^{-3}$); and 
the lower branch connects to the split monopole $C = 0$, $p = 0$
($B_R = \pm 2{\cal P}_0/R^2$). 

Every twisted force-free magnetosphere has a finite toroidal field
\begin{equation}\label{bratio}
{B_\phi(\theta)\over B_\theta(\theta)} = 
\left[{C\over p(1+p)}\right]^{1/2}\,F^{1/p}(\theta).
\end{equation}
A magnetic field line anchored at polar angle $\theta$ will twist 
through a net angle
\begin{equation}\label{deltaphi}
\Delta\phi(\theta) = 2\int_\theta^{\pi/2} 
{B_\phi(\theta)\over B_\theta(\theta)} {d\theta\over\sin\theta}
\end{equation}
before returning to the stellar surface.  Both branches of force-free
equilibria connect to form a one-parameter sequence, labeled by the 
net twist of field lines anchored close to the two magnetic poles,
$\Delta\phi_{\rm N-S} \equiv \Delta\phi(\theta\rightarrow 0)$.  
For modest twists $\Delta\phi_{\rm N-S} \la 1$ one has
\begin{equation}\label{phibrel}
\Delta\phi_{\rm N-S}  \simeq 2\,
\left({B_\phi\over B_\theta}\right)_{\theta=\pi/2}.
\end{equation}

\begin{figure}
\figurenum{2}
\plotone{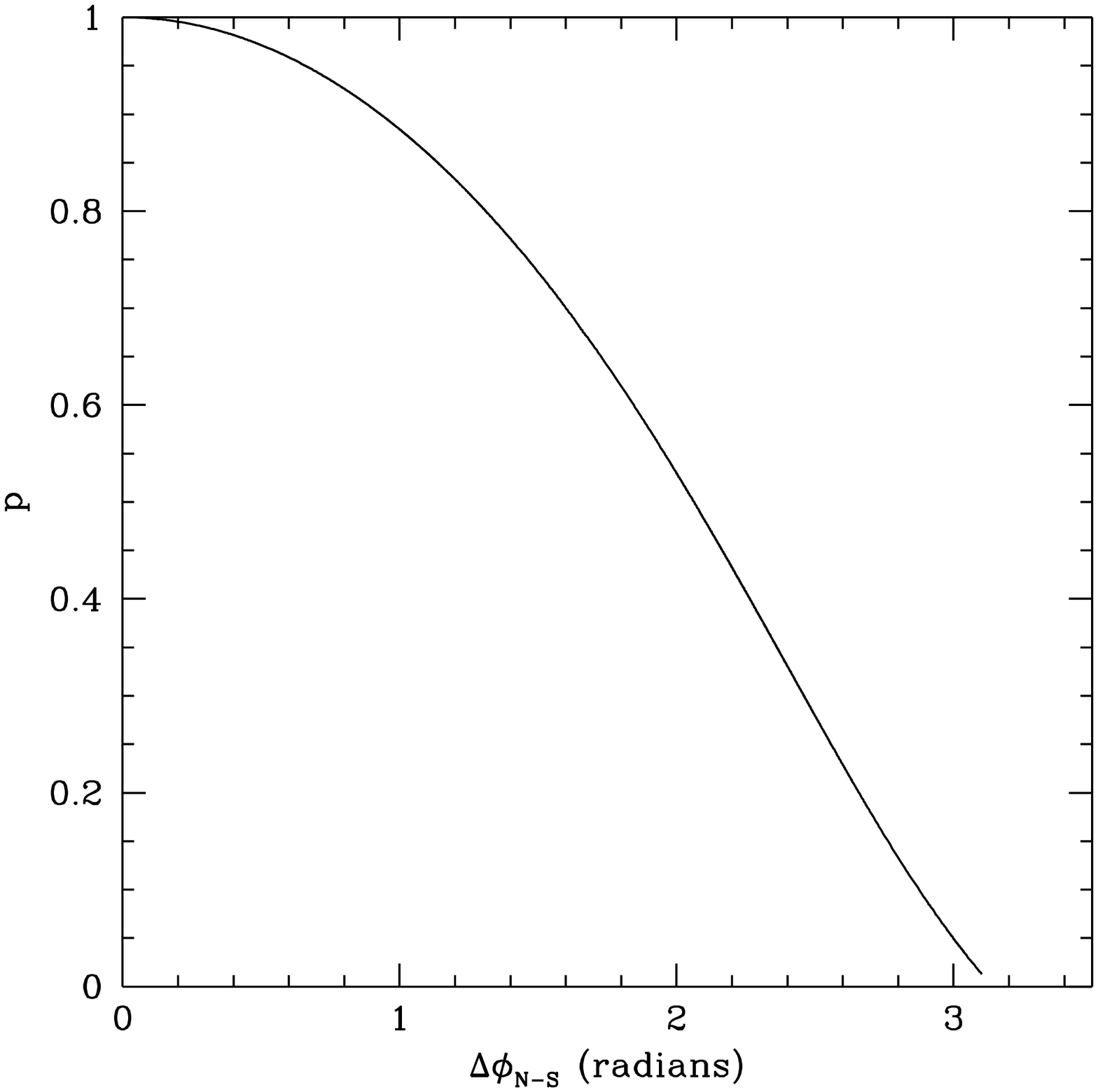}
\caption{The radial index $p$ of the magnetic flux function ${\cal P}$
(eq. [\ref{stream}]) is plotted versus the net twist angle 
$\Delta\phi_{\rm N-S} = \Delta\phi(\theta\rightarrow 0)$ between
the north and south magnetic poles (eq. [\ref{deltaphi}]).
}
\label{pvstwist}
\end{figure}

\begin{figure}
\figurenum{3}
\plottwo{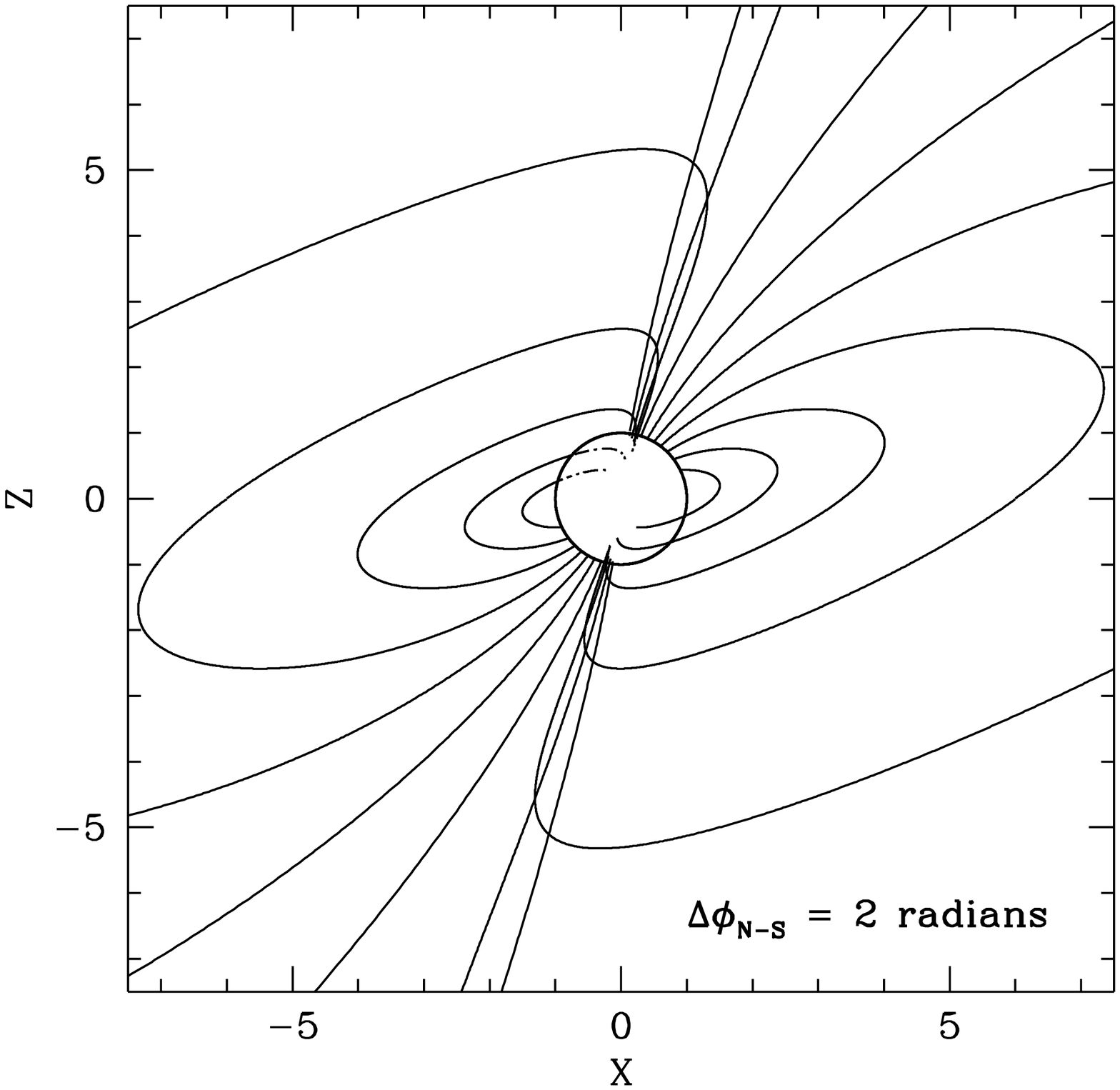}{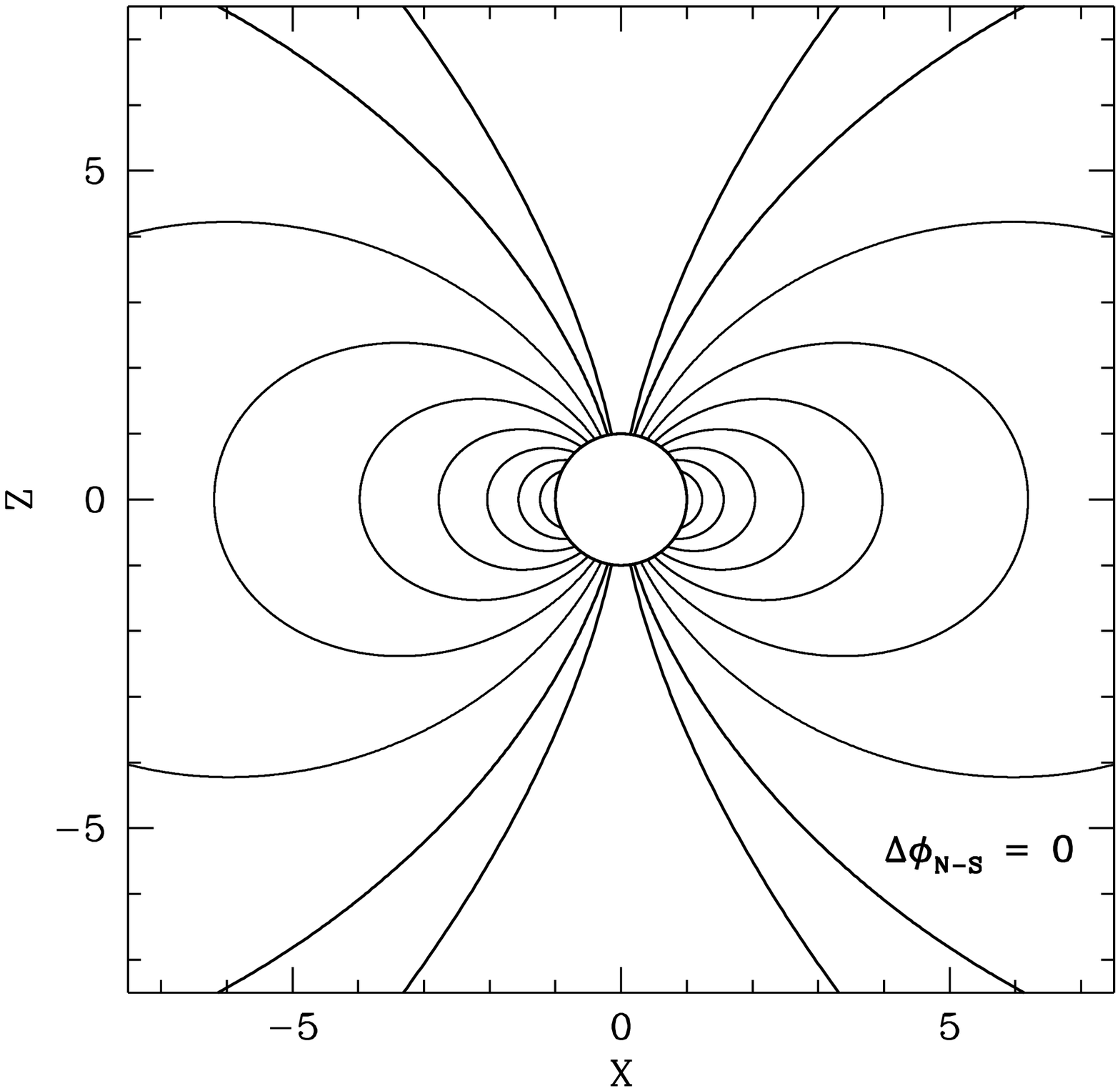}
\caption{An example of a twisted, self-similar, force-free magnetosphere,
with net twist angle $\Delta\phi_{\rm N-S} = 2$ radians.  Only a small
number of field lines are plotted here.  The field lines protruding
from the top right and bottom left corners of the star are anchored in 
the $X-Z$ plane, at regular intervals $\Delta\mu = 0.1$.  Dashed lines
indicate that the field is projected behind the star.  A pure dipole
is shown for comparison.
}
\label{twisted}
\end{figure}

The radial index $p$ is a 
decreasing monotonic function of $\Delta\phi_{\rm N-S}$ (Fig. \ref{pvstwist}).
The radial dependence of the magnetic field softens to
${\bf B} \propto r^{-2.88}$ when $\Delta\phi_{\rm N-S} = 1$ radian.  The net
twist approaches $\Delta\phi_{\rm N-S} = \pi$ (one-half turn) in 
the split monopole limit ($p = 0$).  For comparison, a twisted cylindrically 
symmetric magnetic field pinned in an infinite half-plane expands to infinity
after $1/\sqrt{3}$ turns (Lynden-Bell \& Boily 1994).
One example of a twisted, self-similar magnetosphere (corresponding
to $\Delta\phi_{\rm N-S} = 2$ radians) is depicted in Fig. \ref{twisted}.

These solutions to the force-free equation carry a net helicity
\begin{equation}\label{helicity}
{\cal H}_B = \int {\bf A}\cdot{\bf B}\,dV =
{3\pi\over 2}\,(B_{\rm pole}R_{\rm NS}^2)^2\,\sqrt{C\over p(p+1)}\,
\int_0^1 d\mu {[F(\mu)]^{2+1/p}\over 1-\mu^2}.
\end{equation}
This expression reduces to 
\begin{equation}\label{helicityb}
{\cal H}_B \simeq {2\pi\over 5}\,(B_{\rm pole}R_{\rm NS}^2)^2\,
\Delta\phi_{\rm N-S}
\end{equation}
for modest twists $\Delta\phi_{\rm N-S} \la 1$ radian.  The helicity is
approximately ${\cal H}_B \sim B_\phi B_\theta R^4$, and can
be expressed in terms of the magnetic energy as
${\cal H}_B \sim (B_\phi^2 R^3)\times R$ when the twist is moderately large,
$B_\phi \sim B_\theta$.  This means that the magnetic energy is minimized at
fixed helicity if the twist is concentrated close to the star.  
Indeed, the current density decreases toward the magnetic axis in 
these self-similar solutions (as $J(\theta) \sim \theta^2$ in the 
in the case of a modest twist, $\Delta\phi_{\rm N-S} \la 1$). 

Many further properties of these solutions
are described in Lyutikov \& Thompson (2001).

\subsection{Effects of Slow Rotation}\label{roteffects}

We have constructed self-similar solutions to the force-free equation 
in the infinite volume outside a  nonrotating spherical surface.  A 
real neutron star rotates, and its rigidly co-rotating magnetosphere 
has a finite extent, $R\sin\theta \leq cP/2\pi \equiv R_{\rm lc}$ 
(Goldreich \& Julian 1969).  Close to this speed of light cylinder, 
the rotation will itself cause the field lines to be twisted (Michel 
1991; Mestel \& Pryce 1992) -- but in a different sense than in the 
static, twisted magnetosphere.   Here we discuss some basic effects 
of slow rotation, corresponding to an 
angular velocity of rotation $\Omega \ll c/R_{\rm NS}$.

The rotational sweeping of the magnetic field lines induces an electric field
\begin{equation}\label{efield} 
{\bf E} = -{1\over c}({\bf \Omega}\times{\bf R})\times{\bf B},
\end{equation}
as measured in a background inertial frame.  Here ${\bf\Omega}$ is
the angular velocity of the star.   The component of
${\bf E}$ parallel to ${\bf B}$ is cancelled if the closed field
lines support a net charge density
\begin{equation}\label{chargedens}
\rho = {1\over 4\pi}{\bf\nabla}\cdot{\bf E} =
{1\over 4\pi c}{\bf\Omega}\cdot\left[-2{\bf B} +
{\bf R}\times\left({\bf\nabla}\times{\bf B}\right)\right]
= \rho_{\rm GJ} + \rho_{\rm twist}.
\end{equation}
A second term
\begin{equation}\label{chargedens2}
\rho_{\rm twist} = {1\over 4\pi c}{\bf\Omega}\cdot\left[
{\bf R}\times\left({\bf\nabla}\times{\bf B}\right)\right] \simeq
{1\over c^2}\,{\bf\Omega}\cdot({\bf R}\times{\bf J})
\end{equation}
is now present, as compared with the analysis of Goldreich \& Julian (1969).
(The second equality in this expression applies to field lines which 
close well inside the light cylinder.)  Thus, a current flowing in a 
rotating magnetosphere generates a charge density
\begin{equation}\label{chargedens3}
{\rho_{\rm twist}\over J/c} \sim {R\over R_{\rm lc}},
\end{equation}
near the magnetic equator, decreasing to
\begin{equation}\label{chargedens4}
{\rho_{\rm twist}\over J/c} \sim \left({R\over R_{\rm lc}}\right)^{5/2}
\end{equation}
on the last closed field lines.

The term (\ref{chargedens2}) is easily understood to result 
from the Lorentz boost of the current ${\bf J}$ by the rotational 
velocity ${\bf\Omega}\times{\bf R}$ of the corotating magnetosphere.
In radio pulsar models, it has previously been considered only 
with respect to the open field lines which may carry a space-charge 
limited current from the stellar surface (Mestel 1993).

The density of charge carriers needed to support a static twist
generally exceeds the standard corotation charge density.  At radius $R$
and magnetic co-latitude $\theta$, one has\footnote{We neglect the change
in radial index of the field due to the twist.}
\begin{equation}\label{jvsrhogj}
\left|{J/c\over \rho_{\rm GJ}}\right| \sim \theta^2\,
\left({R\over R_{\rm lc}}\right)^{-1}\,
\left({B_\phi\over B_\theta}\right)_{\theta=\pi/2}.
\end{equation}
Polarization of the positive and negative charges flowing along
the magnetic field can therefore supply a corotation charge
density of either sign.  It is also easy to integrate
the current flowing along field lines which extend out to a radius $R$;
normalizing to the Goldreich-Julian current $I_{\rm GJ} = \Omega^2R_{\rm NS}^3
B_{\rm pole}/2c$, one has
\begin{equation}\label{iratio}
{I(R)\over I_{\rm GJ}} \simeq 
\left({R\over R_{\rm lc}}\right)^{-2}\,
\left({B_\phi\over B_\theta}\right)_{\theta=\pi/2}.
\end{equation}

\subsection{Matching of the Twisted Magnetosphere onto Open Field Lines}
\label{Matching}

On the open field lines, the corotation charge density could be supplied
by a space-charge limited flow of ions and electrons from the surface of 
the neutron star (e.g. Scharlemann, Arons, \& Fawley 1978).  In the absence
of electron-positron pairs, one necessarily has $J \simeq \rho c$ in such
a flow.  Because $\rho_{\rm GJ} = -{\bf \Omega}\cdot {\bf B}/2\pi c$ 
has the same sign above both magnetic poles, a space-charge limited 
flow with $\rho = \rho_{\rm GJ}$ implies a current with an {\it opposing}
sign in the two magnetic hemispheres.  By contrast, the closed loops 
of current flowing through a twisted magnetosphere will maintain a uniform
sign near the magnetic axis.

\begin{figure}
\figurenum{5}
\plotone{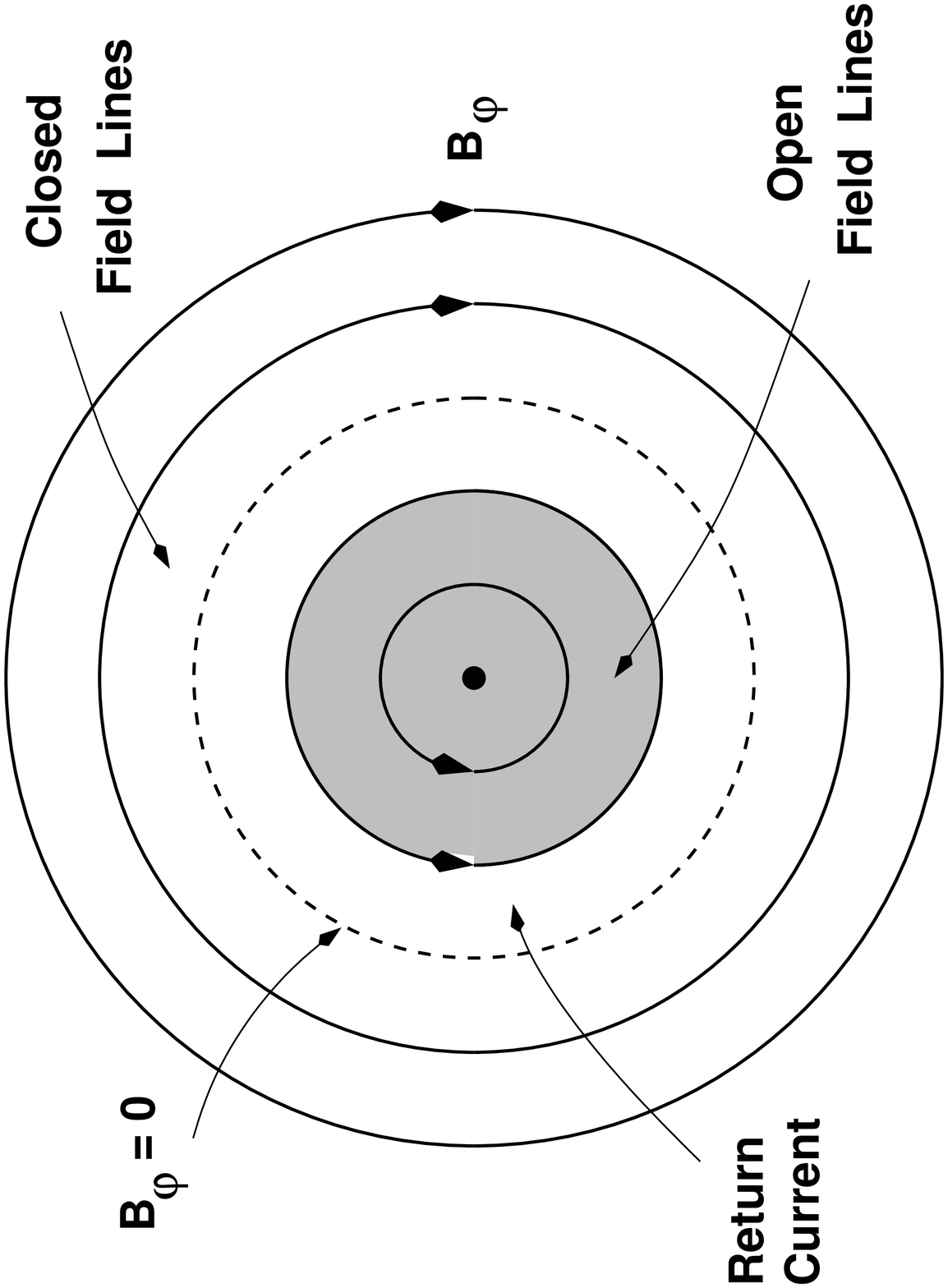}
\label{axialcurrent}
\end{figure}

In the absence of pair creation, such a space-charge limited flow would 
generate, in one hemisphere and close to the magnetic axis,
 a toroidal magnetic field
with the opposite sign to the more global toroidal field.  Figure 
\ref{axialcurrent} depicts 
the distribution of toroidal magnetic field in that hemisphere, assuming 
a geometry for the current flow that is familar from radio pulsar models:
the return current (with a sign opposite to the space-charge limited flow) 
fills a cylindrical sheath near the boundary between open and closed 
field lines.  This return current forces $B_\phi \rightarrow 0$ on some 
cylindrical surface, outside of which $B_\phi$ reverses sign due to the 
global twist (eq. (\ref{iratio})).  

Pair creation in the space-charge limited flow allows an entirely
different relation between charge and current densities $\rho$ and $J$,
because the charges of opposite sign can drift with different speeds:
\begin{equation}\label{rhodec}
\rho = e(n_+ - n_-) \simeq \rho_{\rm GJ},
\end{equation}
whereas 
\begin{equation}\label{jdec}
J = e(n_+v_+ - n_-v_-) \neq \rho_{\rm GJ}c.
\end{equation}
The total number of charge carriers (of either sign) is greatly amplified, 
by a factor $N_{\rm pair} \gg 1$, which we leave as a free parameter.
The net current becomes
\begin{equation}
J \simeq \rho_{\rm GJ}c + N_{\rm pair}|\rho_{\rm GJ}|(v_+-v_-).
\end{equation}
As a result, a small electric field parallel to ${\bf B}$
is sufficient to reverse the sign of $J$ by inducing a small
difference in the speed of positive and negative charges $|v_+-v_-|/c
\sim N_{\rm pair}^{-1}$.
This allows the current to flow in the same direction on the open field lines
as the surrounding closed-field current (and in the {\it opposite} direction
to the original space-charge limited flow).  The
difference in Lorentz factors $|\Gamma_+-\Gamma_-|/\Gamma \sim 
N_{\rm pair}^{-1}$ corresponds to a very modest electrostatic potential 
drop along the outer magnetic field, $|eE_\parallel R| \sim 
(\Gamma/N_{\rm pair})m_ec^2$.  Such a `reversed' current must be intrinsically
time-dependent if the seed charges originate on the surface of the neutron
star:  otherwise, there will be a build-up of charge
above the corotation charge density $\rho_{\rm GJ}$ in the region
where pairs are created.  

The implications of this effect for the observed (narrow) distribution
of spin periods in the SGR and AXP sources are discussed in \S \ref{pnarrow}.
Fast reconnection of these opposing toroidal fields
would force a relaxation of the static twist inside the speed of 
light cylinder, and thus cause a relaxation in the spindown torque.

\section{Resonant Cyclotron Scattering}\label{cycres}

A remarkable feature of these twisted, force-free equilibria is that the
 current currying charges also 
 provide a significant optical depth to resonant cyclotron scattering.
 A particle of charge $Ze$ and 
mass $M$ has a resonance frequency $\omega_c = ZeB/Mc$, and scatters
a photon of frequency $\omega$ (incident at angle $\theta_{kB}$ to ${\bf B}$)
with cross section 
$\sigma_{\rm res}(\omega) = \pi^2 (Z^2e^2/Mc)
\delta(\omega-\omega_c)(1+\cos^2\theta_{kB})$
 (e.g. Canuto, Lodenquai,
\& Ruderman 1971).  In a dipole field, the
resonance sits at a radius 
\begin{equation}\label{resrad}
r_{\rm res} = \left({ZeB_{\rm pole}\over Mc\omega}\right)^{1/3}\,f(\theta) = 
10.5\,Z^{1/3}\,\left({B_{\rm pole}\over 10^{14}~{\rm G}}\right)^{1/3}\,
\left({M\over m_e}\right)^{-1/3}\,\left({\hbar\omega\over 
{\rm keV}}\right)^{-1/3}\,f(\theta),
\end{equation}
in the approximation of static charges.  The angular function
$f(\theta) = (1-{3\over 4}\sin^2\theta)^{1/6}$. 
The importance of resonant scattering may be characterized by the
resonant optical depth $\tau_{\rm res}$:
\begin{equation}\label{taupre}
\tau_{\rm res}(\omega,\theta) = \int n_Z(R)\,
\sigma_{\rm res}\bigl(\omega-\omega_c(R,\theta)\bigr) dR = 
\pi^2 n_Z Ze (1+\cos^2\theta_{kB}){R_{\rm res}\over 
B(R_{\rm res})}\biggl|{d\ln R\over d\ln B}\biggr|,
\end{equation}
where $n_Z$ is the plasma density at the location of the resonance.
The shape of the resonant surface becomes more spherical in the
presence of a net twist, but only slightly when $\Delta\phi_{\rm N-S} 
\la 1$ (Lyutikov \& Thompson 2001).

The plasma density at the resonance may be estimated using the
current density of the self-similar magnetospheric models considered in
Section \ref{force-free}  (eqs. (\ref{ff}) and (\ref{alphaval})),
 $J = (c/4\pi)(B/R_{\rm NS})[C(p+1)/p]^{1/2}
({\cal P}/{\cal P}_0)^{1/p}$.  Assuming 
a drift  velocity  $v^Z \leq c $,
 the
particles of charge $Ze$ and mass $M$ generate an optical depth
satisfying 
\ba
\left({v^Z\over c}\right)\,\tau_{\rm res} &=&
\varepsilon^Z {\pi\over 4}\left(1+\cos^2\theta_{kB}\right)\,
\left[{C(1+p)\over p}\right]^{1/2}\,{[F(\theta)]^{1/p}\over 2+p}
\nn
&=&{\pi\varepsilon^Z\over 4}\left(1+\cos^2\theta_{kB}\right)\,
\left({p+1\over p+2}\right)\,\left[{F(\theta)\over F(\pi/2)}\right]^{1/p}\,
\left({B_\phi\over B_\theta}\right)_{\theta = \pi/2}
\label{tauval}
\ea
(assuming that the particles carry a fraction 
$\varepsilon^Z$ of the current).  
Resonant scattering (at much lower frequencies) by a Goldreich-Julian 
current flowing along open magnetic field lines in a dipolar magnetic field 
has been considered previously by Rajagopal \& Romani (1997).

\begin{figure}
\figurenum{4}
\plotone{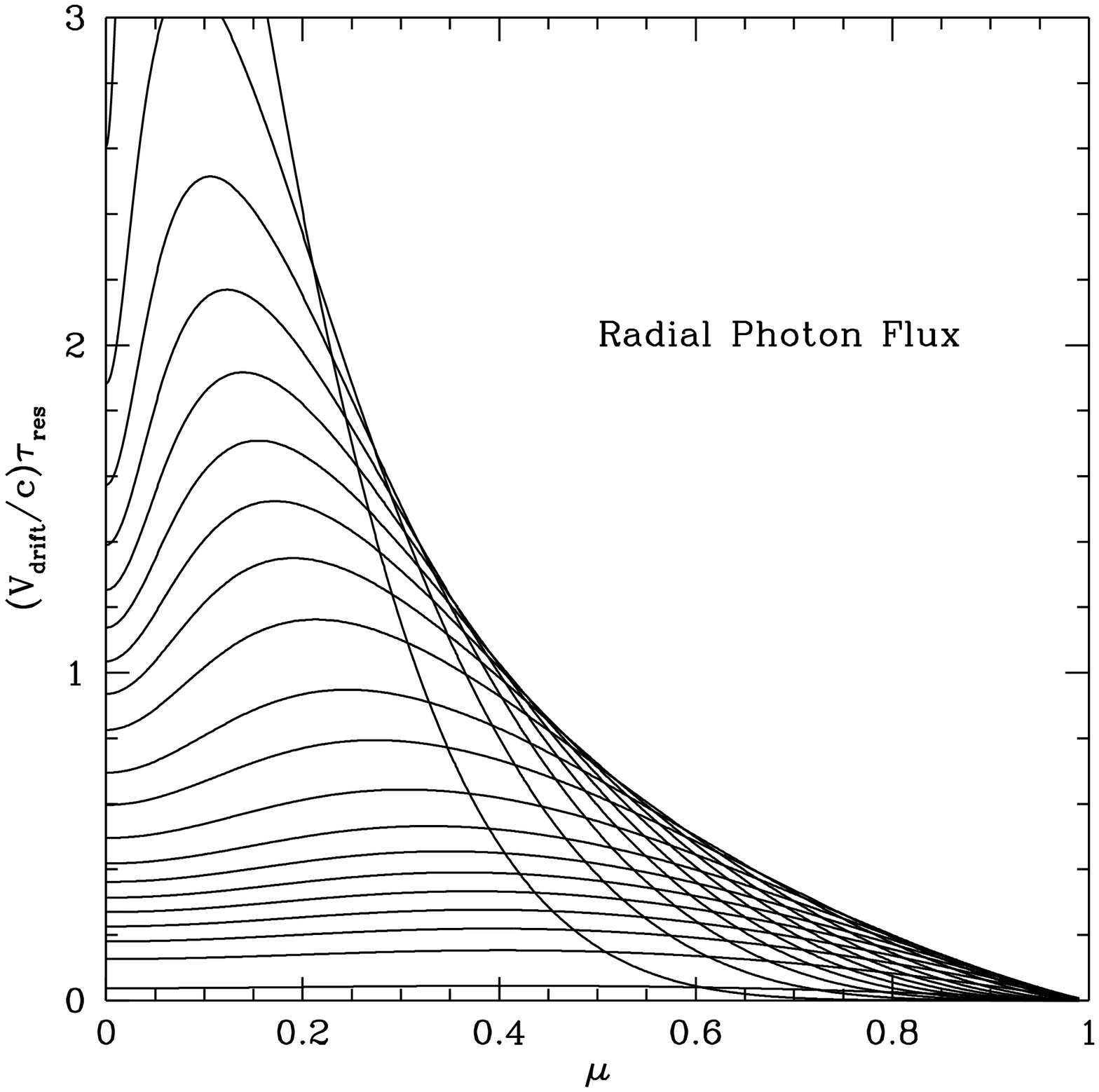}
\caption{Optical depth $\tau_{\rm res}$ to scattering at the 
cyclotron resonance (eq. [\ref{tauval}]) experienced by radially 
streaming photons, as a function of magnetic co-latitude $\mu = \cos\theta$
and the drift speed $V_{\rm drift}$.  This optical depth is calculated
in the approximation that the charges are static.  The various curves 
correspond to different twist angles $\Delta\phi_{\rm N-S}$;  the peak
value of $\tau_{\rm res}(V_{\rm drift}/c)$ is $\sim 0.6$ when the twist 
is $\Delta\phi_{\rm N-S} = 1.5$ radians.  Within a self-similar twisted
magnetosphere, this optical depth is {\it independent} of the radius, 
frequency, or the charge/mass ratio of the scattering particle
(as long as the cyclotron resonance sits outside the spherical surface
in which the magnetic field is anchored).  If the current is supported
by more than one species of charge carrier, as is generally the case,
the plotted value of $\tau_{\rm res}$ must be multiplied by the 
fraction of the current carried by the species of scattering charge.
}
\label{tauprof}
\end{figure}

Two features of the expression (\ref{tauval}) 
 deserve to be emphasized. First, the
product $ (v^Z/c) \tau_{\rm res}$ equilibrates to a value near unity when the
twist is significant, $B_\phi/B_\theta \sim 1$ at the magnetic equator,
{\it independent of the mass and charge of the scatterers, the radius,
or the resonant frequency} (with the obvious proviso that the resonant 
radius sits outside the star).
Second, the optical depth to scattering vanishes near the magnetic
axis, where the equilibrium current density is reduced as the result 
of the large extent of the field lines (Fig. \ref{tauprof}).  
The angular factor in eq. (\ref{tauval}) is 
$[F(\theta)]^{1/p} \simeq \sin^2\theta$ for twists 
$\Delta\phi_{\rm N-S} \la 1$ radian ($F \simeq 1-\mu^2$, $p\simeq 1$). 

Thus the emission  leaving  the surface of the neutron star near 
equatorial plane  will be strongly scattered at the cyclotron resonance, while
at the poles  the emission will emerged almost unscattered.
This axisymmetric, self-similar model will more accurately describe
the magnetospheric structure of an SGR/AXP at larger
distances from the star, where the magnetic field lines are anchored in
a small portion of the neutron star surface, and their arrangement
is less sensitive to irregularities in the surface flux density.
As a result, the scattering surface should more closely approximate 
this simple, axisymmetric form at lower frequencies, and
at the {\it electron} cyclotron resonance than at the ion resonance(s).

A similar result can be derived in any situation where the
magnetic field is strongly sheared, and can be applied to
more complicated (e.g. multipolar) magnetic field geometries.
The resonant optical depth along a ray is inversely proportional 
to the gradient of $B$ parallel to the ray,
\begin{equation}\label{tauresc}
\tau_{\rm res} = \pi^2 (Ze) n_Z (1+\cos^2\theta_{\rm kB}) \,
\left|{dl\over dB}\right|.
\end{equation}
Estimating $(Ze)v^Zn^Z \sim {1\over 2}(c/4\pi)|{\bf\nabla}\times{\bf B}|$
(the resonant charges carry half the current), one has
\begin{equation}\label{tauvalb}
\tau_{\rm res}\left({v^Z\over c}\right) = {\pi\over 8}\,
(1+\cos^2\theta_{\rm kB}) |{\bf\nabla}\times{\bf B}|\,
\left|{dl\over dB}\right|.
\end{equation}
Strong shearing of the magnetic field corresponds to
$|{\bf\nabla}\times{\bf B}|\,\left|{dl/dB}\right| \sim 1$, and 
implies strong Doppler heating within the scattering layer.

\subsection{Equilibrium Drift Speed}\label{vdrift}

Current-carrying charges will feel a drag force when they enter a region
with a significant flux of photons at the cyclotron energy.
In the quiescent state of an SGR (or AXP) source, the drag force acting 
on the ion component of the current is only a modest fraction of gravity
(Appendix B).  The minimum drift speed of the upward moving ions at
the surface of the star can then be related to the gravitational
potential difference which they traverse along a magnetic flux line,
\begin{equation}
\Delta\phi = \left(1-{R_{\rm NS}\over R_{\rm max}}\right)\,
{GM_{\rm NS}\over R_{\rm NS}},
\end{equation}
through 
\begin{equation}
{v_i({\rm min})\over c} = \left({2\Delta\phi\over c^2}\right)^{1/2} 
\simeq 0.6\,\left(1-{R_{\rm NS}\over R_{\rm max}}\right)^{1/2}.
\end{equation}
The drift speed will decrease as the ions climb the potential.

The drag force acting on the electrons
is largest at a radius $R \sim 100$ km, where their
cyclotron energy is in the keV range.  The spectral energy density
of the radiation field can then be estimated as
$U_\omega \sim L_\omega/4\pi R^2 c$.  An electron moving along
${\bf B}$ with a particular speed $v_X$ will feel a Compton drag force
which is perpendicular to ${\bf B}$, so that the electron does not gain
or lose kinetic energy.  In general $v_X$ is a modest fraction of the
speed of light $c$ and does not equal the drift speed $v_e$ of the
current-carrying electrons:  for example, $v_X = \cos\theta_{\rm kB}$ when
the photons stream at a fixed angle $\theta_{\rm kB}$ with respect to
${\bf B}$.  To clarify the discussion that follows, we consider only
the simplest case where the radiation field is directed perpendicular
to ${\bf B}$, and $v_X = 0$.

The component of the Compton draft force parallel to ${\bf B}$ can
be written as
\begin{equation}\label{fdrag}
F_\parallel \sim \Gamma_e^2\,\left({v_e\over c}\right)\,
\int {L_\omega\over 4\pi R^2 c}\,\sigma(\omega,\theta_{\rm
kB}=\pi/2)d\,\omega;
\end{equation}
here $v_e$ and $\Gamma_e$ are the speed and Lorentz of the electron along
${\bf B}$.  Substituting the resonant cross section, one
finds that this drag force is easily sufficient to slow an electron
to sub-relativistic speed, unless it is compensated by an electrostatic
force
\begin{equation}\label{epar}
eE_\parallel \sim {\pi\over 4}\,\Gamma_e^2\,
\left({v_e\over c}\right)\,
\left({e^2\over m_ec^2}\right)\,\left({L_{\omega_c}\over R_{\rm
res}^2}\right).
\end{equation}
The measured spectrum of a magnetar is low enough in the
optical bands (Hulleman et al. 2000) that the ions
feel a negligible Compton drag force at the same radius that the electrons
resonantly scatter off keV photons.  Thus, after exiting the electron
resonance
layer, the ions (which we take to be protons) will have a minimum kinetic
energy
\begin{equation}\label{gamp}
(\Gamma_p-1)m_pc^2 \sim eE_\parallel R_{\rm res}
\end{equation}

Now the assumption of a steady current flow, combined with charge
neutrality ($n_p = n_e$), leads to the requirement that the ratio
of drift speeds
$v_e/v_p$ maintain a constant value $\varepsilon_e/\varepsilon_p$ within
the resonance scattering layer.  (As before, $\varepsilon_i$ is the
fraction of the current carried by particle species $i$.  Notice that
only a tiny charge separation $|n_p-n_e|/n_e = O(10^{-12})$ is needed
to maintain the electrostatic field (\ref{epar}).)  In the simplest
case where $v_e = v_p$, one deduces
\begin{equation}\label{gameq}
{\Gamma_{\rm eq}-1\over \Gamma_{\rm eq}^2(v_{\rm eq}/c)}
\sim {\pi e^2\over 4m_pm_ec^4}\,{(\omega L_\omega)_{\omega_c}\over
R_{\rm res}\omega_c}
\end{equation}
for the equilibrium drift speed (Lorentz factor) within the resonant
scattering layer.  Re-expressing the resonant radius $R_{\rm res}$
in terms of the surface polar (dipole) field, this equation becomes
\begin{equation}\label{gameqb}
{\Gamma_{\rm eq}-1\over \Gamma_{\rm eq}^2(v_{\rm eq}/c)}
\sim 0.9\,
\left[{(\omega\,L_\omega)_{\omega_c}\over 10^{35}~{\rm erg~s^{-1}}}\right]
\,\left({\hbar\omega_c\over{\rm keV}}\right)^{-2/3}\,
\left({B_{\rm pole}\over 10^{14}~{\rm G}}\right)^{-1/3}\,
\left({R_{\rm NS}\over 10~{\rm km}}\right)^{-1}.
\end{equation}
One observes that mildly relativistic motion of the charge carriers
is implied if the X-ray luminosity of the source
is $\sim 10^{35}$ erg s$^{-1}$.
The left side of eq. (\ref{gameqb}) has a maximum value of $\simeq 0.3$ at
$v_{\rm eq}/c \simeq 0.8$.  The current must, as a result, be dominated
by ions when $L_X$ significantly exceeds $\sim 10^{35}$ erg s$^{-1}$.

\section{Implications for X-ray Spectra and Pulse Profiles}

We have described the basic properties (magnetic field geometry,
optical depth to resonant scattering) of a class of twisted, 
self-similar solutions to the force free equation.   In this section,
we show how several observed properties of the SGR and AXP sources
connect directly with this model, which provides a promising framework 
for synthesizing the two populations.  

\subsection{Surface Heating (vs. Volumetric Heating)}\label{surfheat}

Transport of a magnetic field in the deep crust and core of
a magnetar will result in dissipation both inside and outside the
star (TD96).  The heat released in the core (through frictional
heating and beta reactions) is converted largely to neutrinos,
and will be conducted to the surface with a delay of
one year or longer if the surface temperature is $\sim 0.5$ keV.
Non-potential distortions of the magnetic field outside the star 
are converted much more efficiently to visible electromagnetic radiation 
(X-rays).  Changes in the external magnetic field will, as a result, 
induce equally sudden changes in the source brightness (e.g. SGR flares).  

In the present context, where we are interested in a smooth shearing
deformation of the magnetic field, the dissipation outside the star takes
two principal forms:  impact of the current-carrying charges on the
stellar surface;  and resonant Comptonization of this surface X-ray flux
by the magnetospheric currents.  The surface heating provides a
{\it minimal} radiative output of a magnetar, which is increased by
a modest factor as the X-rays undergo multiple scattering (\S \ref{compton}).
When $B_{\rm dipole}$ (multiplied by the irradiated fraction of the
stellar surface) is $\sim 10^{14}$ G, we find that the surface X-ray flux is,
by coincidence, comparable to the equilibrium flux powered by
ambipolar diffusion of a strong internal magnetic field (TD96; Heyl \&
Kulkarni 1998).  The angular pattern of the surface radiation can
be expected to differ significantly from models of passively cooling
magnetars; the implications for pulse fractions are discussed in 
\S \ref{pulses}.

Both positive and negative charges are needed
to cancel the space charge outside the neutron star.  Let
us consider the simplest case where the surface has a light element
composition, and positive ions (e.g. hydrogen, helium, or carbon)
can be emitted thermionically (Thompson et al. 2000).  The
electric field which lifts an ion through the gravitational potential
of the star is many orders of magnitude smaller than the field induced
by a bare space charge $\rho \sim J/c$.  A first estimate of the 
minimum power consumed by the current is obtained by summing the 
gravitational potential energy of the ions falling onto the cathode, 
with the (approximately equal) electrostatic energy picked up by returning 
electrons at the anode, where the ions are accelerated.  The
contribution to the power from an infinitesimal current $J(R_{\rm NS}) dA$
originating in a surface element $dA$
is then $dL_X = 2(GM_{\rm NS} m_p/R_{\rm NS})(1-R_{\rm NS}/R_{\rm max})
J(R_{\rm NS})dA/e$.  Here $R_{\rm max}/R_{\rm NS} = 
[F(\theta)/F(\pi/2)]^{-1/p}$ is the maximum radius attained by a 
field line anchored at polar angle $\theta$.  One observes that
the heating rate is strongly inhomogeneous 
over the surface of the star, even under the assumption of a cylindrically
symmetric and self-similar magnetosphere.   The current density varies
across the stellar surface, as does the gravitational potential shift 
along magnetic flux lines of different lengths. 

Integrating over the surface of the star, we find
\ba\label{lsurf}
L_X & =& {B_{\rm pole} GM_{\rm NS} m_p c\over 2e}\,{p F(\pi/2)\over
2+p}\,\left({B_\phi\over B_\theta}\right)_{\theta = \pi/2}
\nn
&\simeq& 3\times 10^{35}\,\left({B_{\rm pole}\over 10^{14}~{\rm G}}\right)\,
\left({M_{\rm NS}\over 1.4~M_\odot}\right)\,
\left({B_\phi\over B_\theta}\right)_{\theta = \pi/2}
\hskip .3 truein
{\rm erg~s^{-1}}.
\ea
In this self-similar model, the rate of surface heating is close to 
the persistent X-ray luminosity of the SGR and AXP sources 
if $B_{\rm pole}\times (B_\phi/B_\theta)_{\theta = \pi/2} \sim 10^{14}$ G.  
The heating rate is too high if the magnetosphere is strongly twisted 
and $B_{\rm pole} \sim 10^{15}$ G, but in reality the twist could be 
more localized -- e.g. around the magnetic poles.

There is an inefficiency in driving this form of dissipation by
ambipolar diffusion of the core magnetic field:  the total radiative
output of an AXP or SGR is $\sim 10^{47}$ ergs over its $\sim 10^4$
year lifetime, but the total energy radiated in neutrinos is some
$\sim 30$ times higher.  The minimal (r.m.s.) magnetic field needed
to power the observed activity is $\sim 10^{15}$ G, which increases
by a factor $\sim 5$ to allow rapid diffusion of the field through 
the core.\footnote{These fields are easily strong enough to break the
neutron star's rigid crust; but observations of magnetic white dwarfs
demonstrate the existence of stable magnetostatic equilibria even in
perfectly fluid, degenerate stars.  When the magnetic field is pinned
by a rigid crust, transport of the field will force the build-up
of shear stresses in the crust to a (small) fraction of the bulk
magnetostatic stresses; see Thompson \& Duncan (1995), (2001).}
In this regard, the Sun provides a rough analog:  distortions of
the magnetic field outside its photosphere are converted efficiently
to non-thermal radiation (Solar flares).  However, only a tiny fraction
of the total energy stored in the Solar magnetic field is actually
released in its exterior:  only one part in a thousand of the
bolometric Solar output is radiated through the chromosphere and corona.
In a magnetar, motions of the external magnetic footpoints are
driven gradually by diffusive processes in the deep crust and core,
and also suddenly by brittle fractures of its crust.   We focus in
this paper on smooth shear deformations of the external magnetic field,
but the possibility remains that a magnetar will also experience
an approximate analog of `microflaring' on the Sun:  the Hall electric
field will drive high-frequency dislocations of its crust ($\lambda
\la 0.1$ km), which couple efficiently to magnetospheric Alfv\'en waves 
(TD96).

\subsection{Decay of the External Twist}

The preceding result (eq. [\ref{lsurf}]) also allows a simple estimate 
of the maximum time for the external twist to dissipate completely, in 
the absence of further sub-surface motions.  The energy of a twisted 
magnetosphere exceeds the energy of a pure dipole with the same polar 
flux density, 
\begin{equation}\label{emago}
{E_B\over E_B({\rm dipole})} = {3\over 2(p+1)}\int_0^1 d\mu
\left({dF\over d\mu}\right)^2 > 1,
\end{equation}
and can be approximated by 
\begin{equation}\label{emag}
{E_B\over E_B({\rm dipole})} = 1+0.17\Delta\phi_{\rm N-S}^2
\end{equation}
for $\Delta\phi_{\rm N-S} \la 1$.  This works out to
\begin{equation}\label{emagb}
E_B-E_B({\rm dipole}) = 
1.4\times 10^{44}\,\Delta\phi_{\rm N-S}^2\,
\left({B_{\rm pole}\over 10^{14}~{\rm G}}\right)^2\,
\left({R_{\rm NS}\over 10~{\rm km}}\right)^3\;\;\;\;\;\;{\rm  erg},
\end{equation}
for twists $\la 1$ radian.  The decay time for the global twist is then 
\begin{equation}\label{tdecay}
t_{\rm decay} = {E_B-E_B({\rm dipole})\over L_X}
= 40\,\Delta\phi_{\rm N-S}^2\,
\left({L_X\over 10^{35}~{\rm erg~s^{-1}}}\right)^{-1}\,
\left({B_{\rm pole}\over 10^{14}~{\rm G}}\right)^2\,
\left({R_{\rm NS}\over 10~{\rm km}}\right)^3\;\;\;\;\;\;{\rm  yr}.
\end{equation}
very similar to the estimate made in Thompson et al. (2000).

The amount of surface heating would be reduced by a factor
$\gamma_e (m_e/m_p)(c^2/GM_{\rm NS})$ --- and the lifetime of the 
external twist would be increased correspondingly --- if the current 
were carried by relativistic pairs of energy $\gamma_e m_ec^2$.

\subsection{The Ion Cyclotron Resonance and Higher Multipoles}
\label{IonCyclotron}

In this self-similar model, the ion component of the current will generate a 
{\it comparable} optical depth to the electron component in a self-similar
magnetosphere.   The optical depth to resonant scattering
is independent of the net mass or electrical charge, excepting 
the factor $\varepsilon^Z\,(v^Z/c)^{-1}$ (eq. \ref{tauval}.)  
Indeed, the cyclotron energy of an ion of charge $+Ze$ 
and mass $Am_n$ in a magnetic field $B$ is  $\hbar\omega = 
6.3\,(Z/A)(B/10^{15}~{\rm G})$ keV.  Although this resonance sits at X-ray
frequencies in magnetar-strength fields, the resonant radius (\ref{resrad}) is 
$\sim 10$ times smaller for ions than electrons (or positrons).  

The heavier
particles will resonantly scatter a more restricted range of frequencies
(below their surface cyclotron frequency).  This means that higher magnetic 
multipoles will leave a more direct imprint in
the pulse profile through resonant ion scattering, than electron
scattering.  Notice also that higher multipoles will increase the
surface field strength over the polar dipole value, and hence increase
the range of frequencies which resonantly scatter close to the star.  
The pulse profile of the August 27 giant flare provides direct evidence 
for the presence of higher magnetic multipoles in SGR 1900$+$14:
four sub-pulses of a large amplitude appeared during the intermediate 
portion of the burst, which repeated coherently with the 5.16 spin period 
(Feroci et al. 2001; Thompson \& Duncan 2001).  

\subsubsection{Surface Stopping of Magnetospheric Charges
and Ion Cyclotron Emission}\label{cycemisa}

The surface of a magnetar will be impacted by ions or electrons,
depending on the sign of the electrical current.  The characteristic
kinetic energy of an ion is its gravitational binding energy
to the star, $E_i \sim GM_{\rm NS}(A m_p)/R_{\rm NS} = 200\,A$ MeV.
If the ion flow contains a significant proton component, then a layer
of hydrogen will form at the cathode surface.  The downward-moving 
protons will be stopped by $p-p$ collisions at a depth corresponding to
an electron column $\sigma_T N_e \simeq 30$.  

At the anode surface (where the electrostatic
field accelerates the ions of mass $A$ and charge $Z$ upward),
the downward-moving electrons will be pushed to relativistic energies 
$E_e \simeq 200\,(A/Z)$ MeV.  In surface fields $B \ga 10^{15}$ G, 
these particles are not effectively stopped by long-range Coulomb interactions,
because the background electron gas is effectively one-dimensional
(e.g. Nelson, Salpeter, \& Wasserman 1993).  The electrons can be stopped 
by spallation of protons from ions (Thompson et al. 2000),
\begin{equation}
e^- + (Z,A) \rightarrow e^- + (Z-1,A) + p.
\end{equation}
If the anode surface is composed of light ions, e.g. helium or carbon,
then the electrons are stopped at a depth corresponding to\footnote{Using
the cross section of Schaeffer, Reeves, \& Orland (1982) for $e^- + ^{12}C
\rightarrow e^- + ^{11}B + p$.}  $N_e\sigma_T \sim 10^4$.
This stopping mechanism has the interesting consequence that each
electron will create an unbound proton.  If the downward-moving electrons 
and upward-moving ions contribute equally to the current density, 
the proportions of the ion current carried by the protons and the 
heavier (spallated) ions are also approximately equal.

We point out here the interesting possibility that ion cyclotron
emission line may be observed in magnetars whose surfaces are heated
by magnetospheric charges.  
At both ends of the circuit, a significant fraction of the kinetic
energy of the impacting charges is transferred directly to the ions.
This spectral feature is easiest to detect 
if the surface magnetic field lies in the range 
$B_S = (0.2-2)\times 10^{15}(A/Z)$ G, so that the ion cyclotron resonance 
sits between 1 and 10 keV.  Phase-resolved spectroscopy of magnetar 
candidates could, for this reason, provide a direct measure of the 
fraction of the thermal emission powered by magnetospheric currents.

The nature of the radiative transport near the ion cyclotron resonance
is substantially different in `naked' magnetars with persistent
magnetospheric currents, than in accreting neutron stars.  
The density of returning, current-carrying ions is almost constant above 
the surface of a magnetar -- in contrast with an accreting 
neutron star, onto which the infalling matter gradually settles.
The resonant optical depth (eq. [\ref{taupre}]) is proportional to
the local density of scattering charges (rather than to the column density
as with non-resonant scattering).  This means that $\tau_{\rm res}$
has a strong spike just above the surface of an accreting neutron star, 
but remains approximately constant above the surface of an isolated magnetar 
with a non-potential magnetic field.  In the accreting case, the 
result is the formation of an absorption feature at an X-ray cyclotron 
resonance (Zane, Turolla, \& Treves 2000).  In the magnetar case,
by contrast, the kinetic energy lost by the impacting charges will 
raise the temperature in a thin surface layer -- without a large 
cyclotron opacity developing outside that layer.  This allows an 
emission feature to form at a cyclotron resonance.  (Nelson et al. 
1995 have demonstrated this effect for electron cyclotron scattering 
in a $\sim 10^{12}$ G magnetic field, at the surface of a neutron star 
which is impacted by ions moving at the free fall speed.)

\subsection{Resonant Compton Heating}\label{compton}

Multiple scattering at the cyclotron resonance will redistribute photons 
in frequency.  This effect has some similarities with Comptonization in 
an accretion flow (Blandford \& Payne 1981), except that resonant 
scattering of a photon of a given frequency occurs only at a particular 
resonant surface: space and energy diffusion are now directly coupled.
Photons which are backscattered return to a region where their energy
lies below the local cyclotron resonance, and will undergo subsequent
resonant scatterings. 
 
The optical depth to resonant scattering by charges
moving with drift speed $v$ is $\simeq (v/c)^{-1}\,
(B_\phi/B_\theta)_{\theta = \pi/2}$ at the magnetic equator
(eq. [\ref{tauval}]).  Let us first consider the case of
electron scattering.  Resonant scattering by $\sim$ keV 
photons at a radius $R/R_{\rm NS} \sim 10\,
(B_{\rm pole}/10^{14}~{\rm G})^{1/3}$ maintains a mildly relativistic 
electron drift speed when $\omega_c L_{\omega_c} = 10^{35}$ erg s$^{-1}$ 
(\S \ref{vdrift}).  In addition, the product $(v_e/c)\tau_{\rm res}$ has no 
explicit frequency dependence below an energy $\hbar\omega \sim m_ec^2$, and 
above the very low cyclotron frequency at the speed of light cylinder 
$\omega_{c,e}/2\pi \sim 10^8$ Hz.  

In spite of the complicated geometry of the magnetosphere,
we can draw a remarkably simple conclusion:  the product 
of the mean frequency shift per scattering 
$\Delta\omega/\omega$ and the number of scatterings is $O(1)$ when the 
twist angle $\Delta\phi_{N-S} \sim 1$.
As a result, {\it multiple resonant scattering of thermal X-ray photons,
within a self-similar twisted magnetosphere, provides a simple 
mechanism for generating a high energy tail to the X-ray spectrum.}
In the case of electron cyclotron scattering, this high energy tail
will extend up to a frequency $\hbar\omega_{\rm max} \sim {1\over 2}m_ev_e^2 
\ga 100$ keV, where the increase in frequency due to the Doppler effect
is balanced by the Compton recoil.

At a fixed spectral intensity, the hardness of the high energy tail will
increase with the number of scatterings and, hence, with the net twist
imparted to the external magnetic field.  However, this relation
between spectral hardness and magnetospheric current can be 
compensated by changes in spectral intensity, for the reason that the 
drift speed of the electrons varies as a complicated function of
$L_{\omega_c}$ when $(\omega L_\omega)_{\omega_c} \la 10^{35}$ 
erg s$^{-1}$ (\S \ref{vdrift}).  The often measured ratio of power-law 
to thermal intensities should also be interpreted with caution in this 
situation:  if the thermal seed is generated primarily by surface heating,
then its spectrum will not be a pure black body, and a spectral 
decomposition into black body and power-law components will not 
accurately measure the relative luminosities carried by the surface 
and magnetospheric components.

\subsubsection{Scattering at the Ion Cyclotron Resonance}

The optical depths to ion and electron cyclotron scattering 
are comparable in this self-similar model.  More generally,
if the surface magnetic field is strongly sheared, then a 
the product of the resonant optical depth and the ion drift speed 
satisfies $\tau_{\rm res}(v^Z/c) \sim 1$ across a resonant layer
(\S \ref{cycres}).  

However, multiple scattering by ions is different in three 
important respects.  First, as discussed
in Section \ref{IonCyclotron}, it is sensitive to the presence of
higher-order multipoles in the magnetic field.  Second, the
upper cutoff to the non-thermal spectral tail is not set by the 
Compton recoil, but instead by the surface cyclotron frequency,
\begin{equation}
\hbar\omega_{\rm max} = {\hbar\omega_{c,i}(R_{\rm NS})\over
\gamma_i(1-v_i/c)} \simeq 
{30\over \gamma_i(1-v_i/c)}\,\left({Z\over A}\right)\,
\left({B\over 100~B_{\rm QED}}\right)\;\;\;\;\;\;{\rm keV}.
\end{equation}
In this expression, the Doppler factor
$\gamma_i^{-1}(1-v_i/c)^{-1}$ takes into account the (upward) motion
of the ions with speed $v_i$.  Even if the ions are moving 
sub-relativistically, their kinetic energy ${1\over 2} (A m_p) v_i^2$ 
will greatly exceed $\hbar\omega_{c,i}$.  A non-thermal spectrum extending 
above 20 keV can be produced by resonant ion scattering 
if the surface field is stronger than $\sim 3\times 10^{15}$ G, i.e., 
if higher multipoles are present in the magnetic field.  Indeed,
higher multipoles are almost certainly present in the SGR sources
(Feroci et al. 2001).  

Third, a photon has a much larger probability of being backscattered
directly at the neutron star within the ion corona ($R \sim 10-20$ km) than 
in the electron corona ($R \sim 50-100$ km).   Multiple ion
scattering is most effective if ions of different charge/mass ratios
$Z/A$ are present.  For example, protons and heavier ions will carry 
comparable fractions of the current if the electron charge carriers 
returning to the neutron star surface are stopped by spallating protons 
from the ions, as is expected in magnetic fields stronger than 
$\sim 10^{15}$ G (\S \ref{cycemisa}).  The heavier ions, having a lower 
charge/mass ratio $Z/A \simeq {1\over 2}$, will resonantly scatter 
X-rays {\it closer} to the star.  This means that the resonance condition
$\omega = \omega_{c,i}/\gamma_i(1\mp v_i/c)$ can be satisfied for both 
outward-moving and inward-moving photons if the backscattering is by
the protons at a larger radius.   The inner and  outer resonant surfaces 
are converging in a region where the ions are streaming {\it outward} from 
the star (and losing kinetic energy in its gravitational potential)
and an extended high-energy spectral tail will result.

\subsubsection{Comparison with the SGR and AXP Sources.}

The spectra of some AXPs are inferred to have very soft power-law
components (photon index $\sim -4$).  These soft spectra may be explainable
in terms of passive radiative transport through the surface of a neutron star
with a $\sim 10^{14}$ G magnetic field (\"Ozel 2001; Ho \& Lai 2001;
Lai \& Ho 2001).  It should, nonetheless, be kept in mind that the 
spectrum of one AXP is much harder (the photon index is -2.5 for 
1E 1048.1$-$5937; Oosterbroek et al. 1998); and the actively bursting 
SGRs have even harder spectra (photon indices -2.2; Hurley 2000, and 
references therein).  Multiple resonant scattering provides a mechanism 
for generating the full observed range of power-law indices:  the softest 
AXP spectra correspond to external magnetic fields which carry relatively
weak electrical currents, and the hardest SGR spectra to magnetospheres
which are strongly twisted.  

It is interesting to note that $t_{\rm decay}$ (eq. [\ref{tdecay}])
is comparable to the time since the 5 March 1979 giant flare
if the polar field\footnote{To be more precise, the dipole field
in the absence of any external twist.} is $\simeq 10^{14}$ G.
The relatively soft power-law spectrum of SGR 0526$-$66 
(Kulkarni et al. 2001) may therefore have an explanation in 
terms of the decay of the external magnetic shear following the flare. 

As we have discussed, the atmosphere of a magnetar can be expected 
to have more than one scattering corona.  The observations do not 
yet allow us to distinguish between the case where the coronal heating 
is dominated by electron cyclotron scattering, versus ion cyclotron scattering 
(or non-resonant $e^\pm$ scattering) closer to the star.  
Only scattering by electrons can create a non-thermal spectrum
extending up to $\sim m_ec^2$ if the charges are mildly relativistic 
(\S \ref{vdrift}).
For this reason, {\it high energy observations of persistent SGR 
spectra (between 20 and 500 keV) would provide a clear discriminant 
between coronal heating dominated by electron vs. ion scattering.}

More detailed calculations of the distribution of scatterings and
frequency shifts are under way.

\subsection{Pulsations in the Persistent Emission}\label{pulses}

The Anomalous X-ray Pulsars have been proposed to be either magnetars
powered by active field decay (TD96), or passively cooling neutron stars
with $\sim 10^{14}$ G magnetic fields (Heyl \& Hernquist 1997).  
In addition to the constancy of the measured X-ray flux, these models
may also be distinguished by the pattern of the emergent X-ray pulsations.
The X-ray spectra and pulse profiles produced by passively cooling
neutron stars have been calculated by Psaltis, \"Ozel, \& DeDeo (2000)
and \"Ozel (2001), taking into account
the gravitational deflection of the photons, the angular variation of the 
opacity, and the interchange between the two X-ray polarization modes in the
outer atmosphere.  They find that the pulse fractions are less than 
those measured in two AXPs (e.g. Table 2 in Kaspi et al. 2001), even
in the case of cooling through a single, localized hotspot.
However, as we now describe, bombardment of the neutron star surface
by persistent currents, and re-scattering of the X-rays in the magnetosphere,
can lead to significantly different pulse profiles.  

It should first be emphasized that the angular distribution of the
surface X-ray flux depends on the 
relative proportions of the flux carried by the extraordinary 
polarization mode (the E-mode) and the ordinary mode (the O-mode).  
The O-mode radiation is strongly beamed at the surface of a neutron
star, because its scattering opacity scales as\footnote{As before,
$\theta_{kb}$ is the angle between the background magnetic field and
the photon wavevector.} $\sin^2\theta_{kB}$ at frequencies well below the
electron cyclotron frequency (Basko \& Sunyaev 1975).  However, a large
fraction of the energy flux across the surface of a passively cooling
neutron star is carried by the E-mode, whose opacity is more isotropic.

At the surface of a magnetar, the beaming of the O-mode radiation
is most important when the X-ray flux is generated by particle heating.
In that case, most of the surface flux will emerge in the O-mode and,
in addition, the heating may be localized at a small hot spot, which allows
the formation of a narrow cone of radiation.  The opacity of 
the E-mode is suppressed in an ultrastrong magnetic field by a factor 
$(m_e c\omega/eB)^2 =
7\times 10^{-7}\,(\hbar\omega/{\rm keV})^2\,(B/10^{14}~{\rm G})^{-2}$.
Impacting protons will be stopped by $p-p$ collisions at a depth 
corresponding to $N_e\sigma_T \sim 30$ -- well above the position of the 
E-mode photosphere but, generally, below the O-mode photosphere.

\subsubsection{Cyclotron Scattering in the Magnetosphere}

Resonant cyclotron scattering will strongly modify the angular
pattern of the X-ray flux emerging from a twisted, current-carrying 
magnetosphere.  This reprocessing will occur at the electron cyclotron
resonance at a large distance
of $\sim 50-100$ km, where gravitational bending of the photon trajectories 
can be largely neglected.  Reprocessing will also occur at the ion 
cyclotron resonance closer to the star, at $10-20$ km.

Three effects are important here:

1.  The optical depth to resonant scattering is a strong function of
angle in a self-similar, twisted magnetosphere (Fig. \ref{tauprof}).
Even when the twist is large, $\Delta\phi_{\rm N-S} 
\sim 1$ radian, and the optical depth exceeds unity at the magnetic equator,
the resonant scattering surface maintains two holes of solid angle 
$\sim 2$ Sr, centered at the two magnetic poles.  
This effect will increase the relative X-ray flux 
emerging along the magnetic axes.  

2.  The scattered cyclotron radiation is beamed parallel to ${\bf B}$:
$d\sigma_{\rm res}/d\Omega \propto 1+\cos^2\theta$  in the
rest frame of the scattering charge.  A signature of this effect is
the appearance of a fourth-order harmonic in the pulse profile.  Note
also that the degree of beaming can be greatly enhanced by the bulk 
motion of the charge carriers in an electron-ion current.  
Consider, for example, an optically thick scattering screen which 
moves at uniform speed $v$ along the magnetic field.  The intensity of 
the radiation emerging from the screen is concentrated at an angle 
$\theta = \cos^{-1}(v/c)$ with respect to ${\bf B}$.

3.  The resonant surface is aspherical, with a cross-sectional area
that is larger when viewed along the magnetic equator, than from
one of the magnetic poles.  This effect will enhance
the relative flux propagating to infinity along the magnetic equator.

Each of these effects can, by themselves, generate one or two sub-pulses
-- depending on the orientation of the rotation and magnetic axes 
with respect to each other and to the line of sight.  Calculations of 
the resultant pulse profiles are underway.

Even if the magnetic field were axisymmetric in the absence of
electrical currents, deformations of the neutron star crust
which generate the currents could have azimuthal structure.  
Resonant scattering by non-axisymmetric currents therefore provides 
an additional source of structure in the X-ray pulse profile.

\subsection{Implications for Giant Flare Mechanism}

There are two generic possibilities for the production of the
giant flares of the SGRs, in the framework of our model
(Thompson \& Duncan 1995, 2001; Woods et al. 2001a).
First, a giant flare may result from a sudden change (unwinding) in the
internal magnetic field.  In this case, a twist is implanted into
the magnetosphere.  A large-scale displacement of the crust probably 
requires the formation of a propagating fracture, close to which
the magnetic field is strongly sheared.  An attractive mechanism for
powering the flare involves the repeated excitation and relaxation 
of a high current density around the fault.
The energy stored in a twisted internal field, which is available 
for sudden release, can be related to the limiting strain 
$\psi_{\rm max}$ of the crust, 
$\Delta E_B = 1\times 10^{46}\,(\psi_{\rm max}/0.01)^2\,
(B_{\rm pole}/10^{14}~{\rm G})^{-2}$ erg (Thompson \& Duncan 2001).
In the aftermath of the flare, the magnetic field will retain a 
more smoothly distributed component of the shear.
According to this first scenario, the X-ray spectrum should 
be harder when the X-ray flux has returned to the pre-burst value,
but the pulse profile may be simpler
as the result of multiple cyclotron scattering.

Alternatively, giant flares may involve a sudden relaxation in the
twist outside the star, without the impetus of sudden subsurface motions, 
in close analogy with Solar flares.  
This requires that the external magnetic shear build up gradually, and
that the outer crust of the neutron star is deformed plastically by internal
magnetic stresses.  
This mechanism has the advantage that the energy stored in the external 
twist need not be limited by the tensile strength of the crust, but instead 
by the total external magnetic field energy.  From eq. (\ref{emagb}), one
infers that a polar dipole field of $\sim 3\times 10^{14}$ G is needed to 
power a flare of energy $\sim 10^{45}$ erg through the sudden relaxation
of an external twist $\Delta\phi_{\rm N-S}\sim 1$ radian.  
In this second scenario, the X-ray spectrum can be expected to soften 
following the burst, as the magnetosphere becomes more transparent 
to cyclotron scattering.  These differences in post-flare behavior 
may serve to distinguish between the two possibilities.

We now discuss
four pieces of observational evidence which bear upon this question.
Although three seem to support the first hypothesis, the evidence
is mixed.  The most consistent interpretation is, perhaps, that
the giant flares involve a redistribution of current in the 
magnetosphere, which decreases the magnetic shear locally while
at the same time increasing the global twist.

1. The pulse profile of SGR 1900$+$14 changed dramatically following 
the August 27
giant flare, simplifying to a single sinusoidal pulse from 4-5 sub-pulses 
(Woods et al. 2001a).  This change has persisted even after the 
persistent X-ray flux returned to the baseline maintained before the
flare (Woods et al. 2001a).  This decoupling between the pulse profile 
and the total X-ray flux from SGR 1900$+$14, provides evidence that the 
energy source for the persistent emission is concentrated close to the star,
inside the region where the pulse profile is established.  Multiple
scattering of 2-10 keV photons at the electron cyclotron resonance will
cause such a change in pulse profile, if the magnetic field is predominantly
dipolar at a radius of $\sim 100$ km, and is a natural consequence
of a twisting up of the external field during the flare.  By contrast, 
in the second model, the simplified pulse profile would require the elimination
of a non-axisymmetric component of the current during the flare.
This is more difficult to arrange close to the star, where X-rays can
be scattered at the ion cyclotron resonance. 

2.  Both giant flares (on 5 March 1979 and 27 August 1998) were initiated
by a very intense $\sim 0.2-0.5$ s pulse of hard X-rays and gamma rays
(Mazets et al. 1979, 1999; Hurley et al. 1999a; Feroci et al. 1999),
during which the bulk of the flare energy was probably deposited.
This timescale is much longer than the light-crossing time 
of the central magnetosphere (where the energy in the twisted magnetic
field is concentrated);  but is similar to the time for a 
$\sim 10^{15}$ G magnetic field to rearrange material in the 
deep crust and core of a neutron star across several kilometers.
In the second model,
a question of principle also arises as to how a very gradual build up of 
the external shear (over the estimated interval of $\sim 100$ years
between giant flares in any one flare source) could lead to the
sudden release of external magnetic energy on a timescale that is some 
10 orders of magnitude shorter -- without being initiated by a sudden
yield or fracture in a rigid component of the star.

3. The persistent spectrum of SGR 1900$+$14 softened measureably
after the 27 August giant flare: the
best-fit spectral index (pure power law) softened from
-$1.89\pm 0.06$ to -$2.20\pm 0.05$  (Woods et al. 1999a).
(Including a black body component, the best-fit index softened
dramatically from $-1.1 \pm 0.2$ to $-1.8\pm 0.2$.)  This behavior is not
consistent with a twisting up of the outer magnetosphere {\it if} the 
measured spectrum provides a fair measure of the angle-averaged
spectrum and, in addition, if the non-thermal continuum is generated 
by resonant electron cyclotron scattering at $R \sim 50-100$ km.
However, we have seen that regions of strong localized magnetic
shear close to the star provide an alternative location for
the source of this continuum -- through e.g. resonant ion scattering 
(\S \ref{cycres}, \ref{compton}).  The measured spectral change 
in SGR 1900$+$14 may therefore point to a more `compact' scattering corona,
within which the current relaxed even while it increased
on more extended field lines following the flare. 

4.  Both giant flares were followed by multiple repeat bursts of a 
$\sim 1-7$ s duration (Golenetskii et al. 1987; Ibrahim et al. 2001) 
which is intermediate between the flares ($\sim 200-400$ s) and the 
much more common short SGR bursts ($\sim 0.1$ s: Gogus et al. 2001).
These intermediate bursts released less than one percent of the energy
of the giant flares, and most likely represent mild `aftershocks' of
the larger events.  This behavior is possible to describe, in our
axisymmetric magnetospheric model, if the release of energy is gated by the 
rigidity of the crust.   The case of the 29 August 1998
burst from SGR 1900$+$14 -- which had a similar peak luminosity to the
pulsating tail of the August 27 flare but a duration 100 times shorter --
is particularly instructive.  Evidence for a trapped fireball comes 
from both bursts:
from the shape of the declining light curve in the August 27 flare
(Feroci et al. 2001);  and from the presence of an extended faint, 
pulsating tail which 
followed the bright component of the
August 29 burst (Ibrahim et al. 2001).  This faint tail
had a very hard spectrum which softened with time, and can be explained by the
compression and heating of a small patch of the neutron star surface during
the preceding burst.  The high peak flux and short duration of the August 
29 burst require that the magnetically confined fireball had a planar 
geometry, so that it cooled rapidly in one direction without decreasing 
significantly in area.  The simplest interpretation here is that the 
August 29 burst involved a mild slippage of the crust along the same 
fault line which powered the preceding giant flare (Ibrahim et al. 2001).

\subsubsection{Large-Amplitude Pulsations in the Giant Flares}

Within the 27 August 1998 flare itself, the X-ray flux showed pulsations
of a very large amplitude, during all but the first 40 seconds
(Hurley et al. 1999a; Feroci et al. 1999; Mazets et al. 1999).
These pulsations repeated coherently at the 5.16-s spin period of the
star, and the pulse profile maintained a complicated 4-peaked pattern
during the intermediate portion of the burst (Feroci et al. 2001),
which gradually simplified into a single pulse at the very end
(Woods et al. 2001a).  The 25-100 keV light curve recorded by Ulysses
showed a large asymmetry between the first and second halves of the
pulse profile, which was absent in the higher energy 40-700 
BeppoSAX GRBM light curve (Feroci et al. 2001).  The previous
5 March 1979 giant flare also showed large amplitude pulsations
(Mazets et al. 1979).  This narrow collimation of the X-ray flux, combined
with the quasi-thermal flare spectrum, indicates a large angular variation 
in the optical depth.  

The `trapped fireball' model developed in Thompson \& Duncan (1995)
has been successfully tested against the August 27 flare data in 
two respects (Feroci et al. 2001).  After smoothing over the 5.16-s 
pulsations, the flare light curve can be well fit by the contracting 
surface of a magnetically confined $e^\pm$ fireball.  In addition, the 
saturation of the best-fit X-ray temperature at $\sim 12$ keV
during the last $\sim 350$-s of the burst is consistent with the
freeze-out of photon splitting in a Comptonizing atmosphere, as
predicted by Thompson \& Duncan (1995).

The cooling X-ray flux from the surface of a trapped fireball
is concentrated close to the surface of the star, where the magnetic 
scattering opacity is greatly suppressed.  It has been argued that, 
further from the fireball surface, the escaping X-ray flux will 
become collimated along partly open magnetic field lines, as 
the result of the strong inequality in the scattering opacity 
of the two X-ray polarization modes in a super-strong magnetic field 
(Thompson \& Duncan 1995, 2001).  In particular, the opacity of the
E-mode scales as $\sim B^{-2}$ and grows rapidly with radius.  Thus,
matter suspended higher in the magnetosphere by the hyper-Eddington
flux can provide a dense scattering screen that is optically thick
to {\it Thomson} scattering, and through which the X-rays can escape 
only by pushing the matter to the side.  The burst light curve therefore
provides information about the connectivity of the magnetic 
field lines close to the fireball surface.  As the fireball shrinks, 
it connects with a smaller (and therefore more regular) portion of the 
magnetosphere -- which could explain the reduced number of sub-pulses 
toward the end of the burst.  

In the self-similar magnetospheric model which we have described, 
the resonant scattering opacity varies too smoothly with angle to explain the 
narrow collimation  of the  X-ray flux observed in both giant flares.
Even during a giant flare, the photon flux is too weak to break open
the closed field lines at a radius of $\sim 50-100$ km
(the position of the electron cyclotron resonance of a $\sim 40$ keV
photon).  The energy density in the magnetic field 
exceeds that in the freely streaming X-rays by a large factor $\sim 
10^6\,(L_X/10^{42}~{\rm erg~s^{-1}})^{-1}\,(\hbar\omega/40~{\rm keV})^{4/3}\,
(B_{\rm pole}/10^{14}~{\rm G})^{2/3}$ at the resonant surface,
and so the radiation pressure imparts only a negligible distortion 
to the field lines.  Thus, we must consider whether a sudden
increase in  the magnetospheric current during the first $\sim 1$ second
of the flare (when most of the flare energy was probably released
into the magnetosphere) could force the multiple X-ray beams to be 
re-scattered at the electron cyclotron resonance, and thereby isotropized. 

During an outburst, the current-carrying electrons feel an enormous 
drag force where their cyclotron energy lies in the X-ray range, 
independent of the sign of the current (Appendix \ref{compforce}).  In the
quiescent state, the electrons move in opposite directions
(both toward the neutron star and away from it) on different portions 
of a closed current loop.  However, charged particles may be injected
into the magnetosphere at a much higher rate during an SGR outburst,
than they need be to support the current associated with a static twist.
The quiescent charge flow is given by eq. (\ref{iratio}), and can
be converted to a kinetic luminosity 
\begin{equation}\label{lmatter}
L_{\rm matter} \sim {I(R)\over e} m_p c^2
= 1\times 10^{35}\,\left({B_{\rm pole}\over 10^{14}~{\rm G}}\right)^{1/3}\,
\left({\hbar\omega\over {\rm keV}}\right)^{2/3}\,\left({R_{\rm NS}\over
10~{\rm km}}\right)\;\;\;\;\;\;{\rm erg~s^{-1}}.
\end{equation}
(Here we have expressed the resonant radius in terms of the cyclotron 
photon energy.)   This luminosity is comparable to the persistent 
X-ray luminosity (see eq. [\ref{lsurf}]), but is less than one part 
in a million of the bursting luminosity.  
In a giant flare, the radiation pressure is high enough to advect 
matter outward at relativistic speed from the heated neutron star 
surface (Appendix \ref{compforce};  Ibrahim et al. 2001).
Only a minuscule fraction of the bursting luminosity need be carried 
by this entrained matter, in order to supply the charges which
support a static current flowing along the twisted magnetic field.  

The current can be maintained by a small differential
drift of the ions with respect to the electrons and photons,
in the presence of a dense wind from the neutron star surface.
The cyclotron energy of the ions is too small to couple them tightly 
to the X-rays at the same radius where the electrons are tightly coupled.
This allows the X-rays to flow outward, largely unimpeded, independent of
the sign of the current.  However, because the electrons are tightly
coupled to the photons, a multipolar pattern can be maintained in 
the X-ray flux {\it only if the dense matter transported to large radius 
reflects the multipolar pattern of the magnetic field close to the source.}

It should also be noted that the pulse profile is measured in a much 
higher energy range during the flare (25-100 keV for Ulysses and 40-700 
for BeppoSAX) than it is in quiescence (typically 2-10 keV).  As a result, the 
pulse profile is more sensitive to the presence of higher magnetic multipoles
during outburst, than in quiescence.

\subsubsection{Implications of the Variable Quiescent Pulse Profile of 
SGR 1900$+$14}

The 2-10 keV pulse profile of SGR 1900$+$14 was complicated and multi-peaked 
before the 27 August 1998 giant flare (Hurley et al. 1999c).  
Our explanation for the change in the pulse profile following the flare 
(to a single sinusoidal pulse) involves a re-scattering of the X-rays 
at the electron-cyclotron resonance.  If during the flare
an additional twist was implanted in the magnetosphere, then there
was at the same time an increase in the current density and the
optical depth to resonant scattering.

The smooth, single pulse observed following the flare implies a 
simple -- predominantly dipolar -- geometry of the poloidal magnetic 
field at a distance of $\sim 50-100$ km from the neutron star.
We can then derive an upper bound to the strength of the surface quadrupole.

Neglecting departures from axisymmetry and reflection symmetry, 
the poloidal field can be decomposed as
\begin{equation}\label{bequator}
B_\theta(R,\theta = \pi/2) = \sum_{\ell=1}^\infty B_\ell(R_{\rm NS})\,
\left({R\over R_{\rm NS}}\right)^{-(\ell+2)}
\end{equation}
at the magnetic equator.
Here, $B_1 = {1\over 2}B_{\rm pole}$, and higher $\ell$ represent 
higher multipoles.  We require that the quadrupole $\ell = 2$ remain 
weaker than the dipole $\ell = 1$  at the electron cyclotron 
resonance of a 10 keV photon.  Expressing the resonant radius as 
$R_{\rm res}/R_{\rm NS} = (B_1/B_{\rm QED})^{1/3}\,
(\hbar\omega/{\rm keV})^{-1/3}$, one deduces
\begin{equation}\label{bquadb}
{B_2(R_{\rm NS})\over B_1(R_{\rm NS})} \la 5\,
\left[{B_1(R_{\rm NS})\over 10^{14}~{\rm G}}\right]^{1/3}\,
\,\left({\hbar\omega\over 10~{\rm keV}}\right)^{-1/3}.
\end{equation}

The complicated 2-10 keV pulse profile observed before the August 27 flare then
has two explanations in our model.  The first possibility is
that, even before the flare, the X-ray flux was reprocessed
by resonant scattering off electrons at $R \sim 50-100$ km,  but that
the current was not axisymmetric.
The second possibility is that the current flowing along extended field 
lines was small, so that the complicated pulse profile resulted from
anisotropic emission and scattering close to the source.  For example,
the ion cyclotron resonance lies in the X-ray range near the
surface of a magnetar, where the field is probably dominated by higher 
multipoles (\S \ref{IonCyclotron}).

These two models lead to differing conclusions about the overall strength of 
the magnetic field at the surface of SGR 1900$+$14.  Polar electrical currents 
will accelerate the rate of spindown with respect to a simple 
magnetic dipole, and reduce the surface field inferred 
from the measured period and period derivative (see \S \ref{spindown}).  The
rapid spindown of the SGRs 1900$+$14 and 1806$-$20 compared with the 
AXPs, and the displacement of SGR 1900$+$14 from the center of the nearest
supernova remnant, suggests that their spindown is {\it transiently}
accelerated (Thompson et al. 2000).  Such a temporary increase in
the rate of spindown could be effected through a global twist 
imparted to the magnetosphere.

In the absence of a global twist, SGR 1900$+$14 is inferred 
to have a polar dipole magnetic field $B_{\rm MDR} \sim 2\times 10^{15}$ G,
based on the rate of spindown before the giant flare (Kouveliotou et al. 
1999).  The presence of higher magnetic multipoles (needed to explain 
the complicated angular pattern of resonant ion scattering) then 
guarantees yet stronger surface fields, and allows multiple ion scattering 
to occur up to a high energy cutoff of 
$\sim 40\,(B_{\rm surface}/3B_{\rm MDR})$ keV. 

A check of these ideas is provided by a relatively short 3.5 s burst 
emitted by SGR 1900$+$14 on 29 August 1998, two days after the giant flare.
This burst was followed by a faint ($L_X < 10^{37}$ erg s$^{-1}$) and 
very extended tail, which lasted more than 1000 s and gradually merged 
with the persistent emission (Ibrahim et al. 2001).  The 2-20 keV pulse 
profile in this tail had a similar shape, and maintained a constant relative 
phase, with the later persistent emission (Palmer 2001).  Even at its 
peak, the radiative flux of the tail did not exceed that of the persistent 
emission\footnote{The August 29 burst itself occurred during a period
of enhanced persistent emission following the August 27 giant flare
(Woods et al. 2001a); this comparison is made with the amplitude of 
the persistent emission recorded just before the August 29 burst.} 
by more than an order of magnitude.  Thus, the pulse profile should
be controlled by the same resonant scattering processes in the decaying
tail, as it is in the later persistent emission, and the observed
constant phase alignment of the pulse is expected.

\section{Implications for SGR/AXP spindown}
\label{spindown}

The measured spindown of SGR 1806$-$20 (Kouveliotou et al. 1998)
and SGR 1900$+$14 (Kouveliotou et al. 1999; Woods et al. 1999c; Marsden
et al. 1999) corresponds to a polar dipole field $B_{\rm pole}
\simeq 2\times 10^{15}$ G.  In the presence of a net twist, the
external magnetic field drops off more slowly than $\sim R^{-3}$
(Fig. \ref{pvstwist}).  The field strength $B(R_{\rm lc})$ at the speed 
of light cylinder $R_{\rm lc} = cP/2\pi$ is then stronger than a pure dipole,
$B_\theta(R_{\rm lc})/B_{\rm pole} =  {1\over 2}pF(0)\,
(R_{\rm lc}/R_{\rm NS})^{-(2+p)}$.
Since the rate of loss of rotational energy is $I\Omega\dot\Omega
\sim B_\theta(R_{\rm lc})^2 R_{\rm lc}^2 c$, the flaring of the
field causes the spindown rate to increase. Equivalently, the
real
polar surface field {\it decreases} from
the one  inferred from a measured $\dot P$ and $P$
with increasing twist (decreasing radial index $p$).  Compared directly
with the magnetic dipole value, it is
\begin{equation}\label{bcompare}
{B_{\rm pole}\over B_{\rm pole}(p = 1)} = {1\over pF(0)}
\left({cP\over 2\pi R_{\rm NS}}\right)^{p-1}.
\end{equation}
Notice that there is no direct dependence on $\dot P$ in this expression.
An additional consequence is to reduce the braking index below the 
dipole value,
\begin{equation}\label{brake}
n = {\ddot\Omega\Omega\over (\dot\Omega)^2} = 2p+1.
\end{equation}

\begin{figure}
\figurenum{6}
\plotone{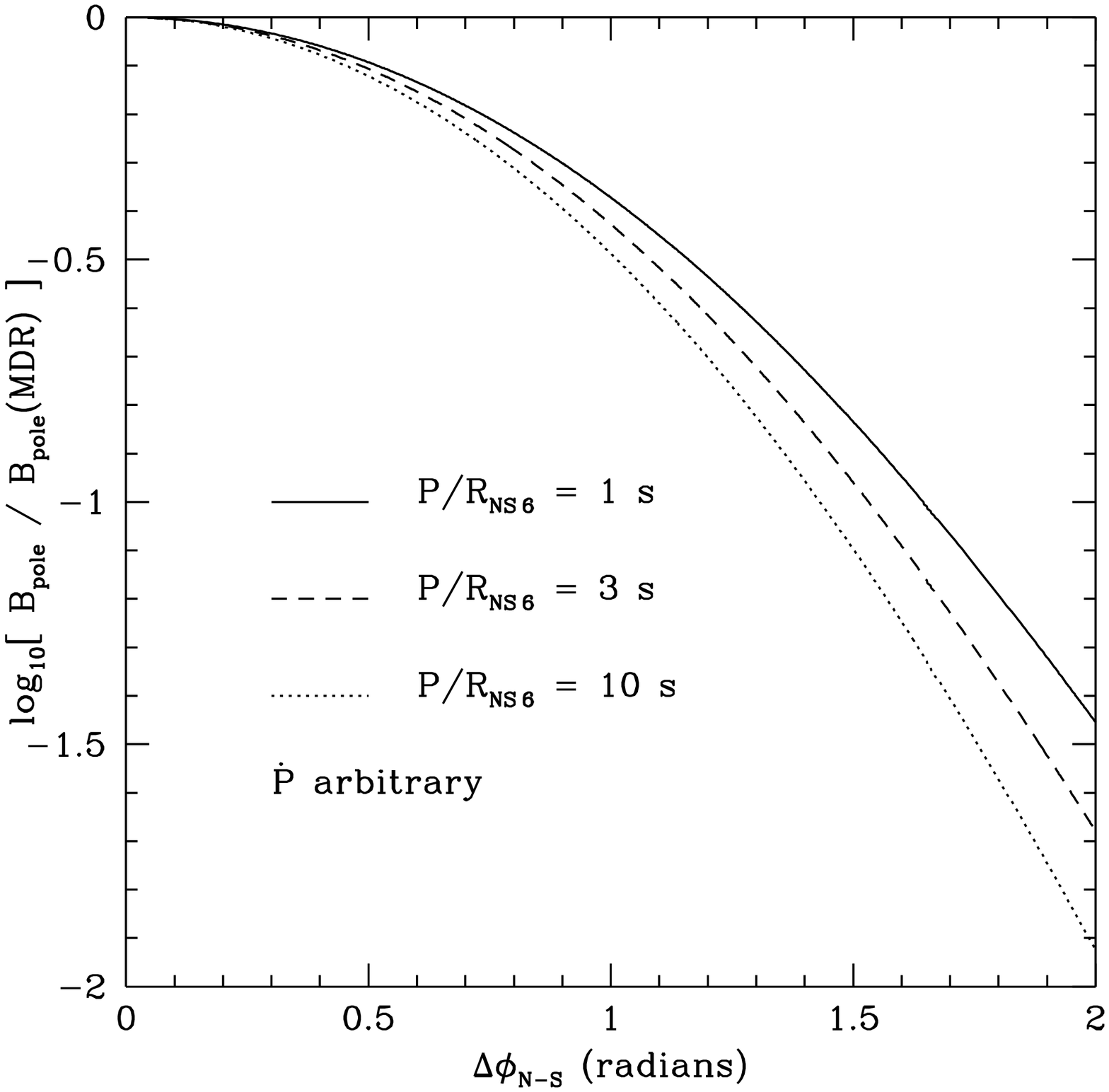}
\caption{Polar surface magnetic field, as inferred from a measured
spin period $P$ and period derivative $\dot P$, compared with the
classical magnetic dipole formula (eq. [\ref{bcompare}]) evaluated
at $\sin\alpha = 1$.
}
\label{bsurface}
\end{figure}

The ratio (\ref{bcompare}) is plotted in Fig. \ref{bsurface} 
for various spin periods.
The true polar field is smaller by a factor $\simeq {1\over 3}$
when the outer magnetosphere has a twist $\Delta\phi_{\rm N-S} = 1$ radian,
for spin periods in the range $3-10$ s;  whereas a twist 
$\Delta\phi_{\rm N-S} = 1.5$ radian leads to a reduction of one
order of magnitude in $B_{\rm pole}$.  This model has the further
implication that, for a fixed polar field, {\it the spindown rate increases
with the optical depth to resonant scattering, and hence with the
hardness of the persistent X-ray spectrum}.   Indeed, the active SGRs
1806$-$20 and 1900$+$14 both have higher $\dot P$ and harder X-ray spectra
than any AXP -- a trend which has been further quantified by
Marsden and White (2001) for the combined population of SGR and AXP
sources.  The quiescent SGR 0526$-$66 has a softer
spectrum (Kulkarni et al. 2001).  We predict that its spindown rate,
when measured, will be intermediate between these sources and 
the AXPs.  

It should be emphasized that if the spindown of a magnetar is
{\it persistently} accelerated in this manner (so that the magnetospheric
twist remains constant), then its characteristic age hardly differs from
the magnetic dipole value.   The spin frequency decreases as $\Omega(t)
\propto t^{-2p}$, and so the characteristic age is larger than
$P/2\dot P$ by a factor $1/p$.  In the case
of the AXP 1E 1841$-$045, the near equality between the $4\times 10^3$ yr
characteristic age and the age of the surrounding
Kes 73 (Gotthelf et al. 1999) {\it does not} imply that its 
magnetic field must be close to a true dipole.

It has been previously noted (Thompson \& Blaes 1998; Harding,
Contopoulos, \& Kazanas 1999; Thompson et al. 2000) that 
persistent seismic activity in a magnetar
can also increase the rate of spindown with respect to a vacuum magnetic
dipole -- by triggering a fluctuating current in the magnetosphere
which drives an outward flux of particles and Alfv\'en waves.
The real polar field 
of SGRs 1806$-$20 and 1900$+$14 is reduced by a factor $\sim 3$ with respect 
to the  one inferred from 
to the magnetic dipole formula, if the persistent seismic luminosity
is equal to the observed X-ray luminosity of $\sim 10^{35}$ erg s$^{-1}$.
The plerionic synchrotron emission powered by a persistent
particle wind of luminosity $\sim 10^{35}$ erg s$^{-1}$ could,
in fact, have escaped detection.
Measurements of spindown in SGR 1900$+$14 or SGR 1806$-$20 do not, however, 
show a direct correlation between the rate of spindown and bursting activity --
with the noticeable exception of the August 27 giant flare itself
(Woods et al. 1999c, 2001b).  For this reason, static magnetospheric currents 
seem a more promising source of non-uniform spindown in the SGR and AXP 
sources.  

Ejection of a large number of particles during a giant flare would cause 
a transient spindown of a soft gamma repeater.  As we now show, the cumulative
torque is larger if the external field is twisted, than if it is
dipolar close to the star.   The spin period of SGR 1900+14 did indeed
increase by $\Delta P/P = 1\times 10^{-4}$ (in comparison with the
extrapolation of the previously measured spindown) within
three months of the 27 August 1998 giant flare (Woods et al. 1999a).
However, the torque calculated assuming a dipolar (near) field is too small by
an order of magnitude if $B_{\rm pole} \sim 10\,B_{QED} = 4.4\times 10^{14}$
G, and if the particle energy $\Delta E$ and duration $\Delta t$ of
the outflow are normalized to the energy and duration of the X-ray
outburst (Thompson et al. 2000).  Estimating
\begin{equation}
I{\Delta\Omega_{\rm NS}\over\Omega_{\rm NS}} \simeq 
-{2\over 3c^2}\Delta E R_A^2,
\end{equation}
where the Alfv\'en radius is determined by balancing the ram pressure
of the particles with the magnetic tension,
\begin{equation}
{\Delta E/\Delta t\over 4\pi R_A^2 c} = {B^2(R_A)\over 4\pi}
= {B_{\rm pole}^2\over 4\pi}\,\left({R_A\over R_{\rm NS}}\right)^{-2(2+p)},
\end{equation}
one finds that the net torque is increased by a factor
\begin{equation}\label{deltapp}
{(\Delta P/P)_{p = 0.8}\over (\Delta P/P)_{p = 1}} = 
4\,\left({B_{\rm NS}\over 10\,B_{\rm QED}}\right)^{1/9}\,
\left({\Delta E\over 10^{44}~{\rm erg}}\right)^{-1/18}\,
\left({\Delta t\over 400~{\rm s}}\right)^{1/18}
\end{equation}
when $p$ is reduced from 1 to 0.8.  This brings the calculated
torque close to the observed value if $B_{\rm pole} \sim 10\,B_{\rm QED}$.

\subsection{Narrow Distribution of SGR/AXP Spin Rates}\label{pnarrow}

The SGR sources
have spin periods measured\footnote{Derived from the persistent emission
for SGR 1806$-$20 (Kouveliotou et al. 1998), from the persistent emission
and 27 August 1998 giant flare for SGR 1900$+$14 (Hurley et al. 1999a,b;
Feroci et al. 1999; Mazets et al. 1999), and from the 5 March 1979
giant flare for SGR 0526$-$66 (Mazets et al. 1979).} in the range 5-8 s.
The distribution of spin periods for the AXP sources is remarkably
similar:  6-12 s (e.g. Mereghetti 2000).  While the number of sources
is small enough that the detection of much shorter spin periods
is not surprising, even if the sources are born spinning much more rapidly,
it {\it is} surprisingly narrow given the wide range of characteristic
ages $P/\dot P$  -- from $1-3\times 10^3$ yr for SGRs 1806$-$20 and 1900$+$14
up to $4\times 10^5$ yr for the AXP 1E 2259$+$586.  It has been suggested,
as a result, that the spindown of the active SGRs is transiently
accelerated, and that the spindown of 1E 2259$+$586 (which sits near 
the center of the much younger SNR CTB 109) has decayed significantly from
its long term average (Thompson et al. 2000).    Clearly transient
acceleration is possible in this model if the AXPs are (mostly) dormant
SGRs, and the external magnetic field is twisted up during periods of 
burst activity.

The narrow range of 
spin periods is suggestive of some physical process which limits 
the spindown rate beyond a period of $\sim 8$ seconds.  We would like
to point out that this period is remarkably close to the upper
envelope of the distribution of spin periods in the known radio pulsar 
population.  

The magnetospheric model which we have outlined 
provides a motivation for a reduction in torque above a critical
spin period, where the potential drop through the magnetosphere is
no longer high enough to trigger a pair cascade through emission
of curvature $\gamma$-rays.  In the absence of pair creation, the 
space-charge limited flow along open magnetic field lines will generate, 
in one hemisphere, a toroidal magnetic close to the magnetic axis with 
the opposite sign to the more global toroidal field (Fig. \ref{axialcurrent}).
Fast reconnection of these opposing toroidal fields
would have the effect of forcing a relaxation
of the static twist inside the speed of light cylinder.
In other words:  beyond the pair death line, the intermediate regions of the
corotating magnetosphere probably cannot maintain a static twist.
Current will diffuse away from the region closest to the star (where
the most of the current is concentrated in our self-similar
solutions: eq. [\ref{iratio}]) only on the relatively 
long timescale (\ref{tdecay}).   

We should emphasize that we have not yet been able to 
demonstrate unambiguously the opposite effect:  that pair creation
will act to stabilize a large-scale twist.  Pair creation does cause
a huge multiplication in the number of charges flowing outward on 
open field lines, so that the pair-loaded plasma is capable of maintaining 
a net Goldreich-Julian charge density -- even while the axial current flows 
in the {\it opposite} direction to the original space-charge limited flow 
(\S \ref{roteffects}).   Whether the global current flow actually 
takes advantage of this effect is a question that we cannot presently
address from first principles.

\section{Summary}

We have shown that several properties of magnetar candidates in their
quiescent states become easier to understand if the neutron star's magnetic
field is {\it globally twisted}.  These properties are directly affected
in a {\it correlated} manner by persistent magnetospheric currents.
The observed X-ray pulse profile and spectrum are modified by
resonant cyclotron scattering and the spindown torque is increased
(for a fixed surface field strength) over the standard vacuum dipole formula.

We have idealized the magnetosphere as a twisted dipole, and
constructed self-similar solutions to the force-free equation.
The self-similar ansatz requires that the surface flux density and
the current density (related to the twist of the field lines) have
a particular dependence on the magnetic latitude which changes shape
depending on the strength of the current.  As the net twist between
the north and south magnetic hemispheres $\Delta\phi_{\rm N-S}$
increases from 0 to $\pi$, the external field continuously interpolates
between a dipole and a twisted, split monopole.  Since axisymmetric
shear deformations of an axisymmetric star do not change the angular
distribution of $B_R$, actual deformations of a magnetar must
be represented by some non-linear combination of these solutions.
This also suggests that the magnetosphere of a neutron star may, 
in practice, not be able to maintain a twist larger than $\Delta\phi_{\rm N-S}
\sim 1-1.5$ radians (above which the distribution of flux with
polar angle begins to differ significantly from a pure dipole).

Our principal conclusions can be summarized as follows:

1. It has previously seemed difficult to make deductions about 
the configuration of the magnetic field in the SGR and AXP sources:  
one obtains evidence
for the presence of higher multipoles from the light curves of the giant
flares (Feroci et al. 2001; Thompson \& Duncan 2001) but not much more than
that.  The existence of magnetars was motivated by considerations of
magnetic field amplification through dynamo activity in young, convective
neutron stars:  a large-scale helical dynamo is possible when the
initial spin period is shorter than $\sim 3$ msec (the convective
overturn time of nuclear matter from which neutrinos are escaping
with a luminosity $L_\nu \ga 10^{52}$ erg s$^{-1}$) (Duncan \& Thompson
1992; Thompson \& Duncan 1993).  Nonetheless, the connection between
this theoretical dynamo model and the (now) slowly rotating SGR and AXP
sources has seemed tenuous.  The new results presented in this paper --
which indicate the presence of strong internal toroidal fields in
the SGR and AXPs -- show that this theoretical picture has, at least,
a degree of self-consistency.

2. A persistent current can be maintained by electrons and ions stripped
from the neutron star surface, if the elemental composition is light 
(e.g. hydrogen, helium, or carbon).  The impact of the returning ions at 
the cathode region of the neutron star surface, and the downward acceleration
of the returning electrons at the anode surface (where the ions are
accelerated upward) generates a luminosity $L_X \sim
10^{35}\,(B_{\rm pole}/10^{14}~{\rm G})$ erg s$^{-1}$ in surface
X-ray emission if the entire magnetosphere is twisted.  This luminosity 
provides a {\it lower bound} to the electromagnetic output of a 
neutron star with an actively decaying magneic field and is, by 
coincidence, comparable to the passive X-ray flux powered by ambipolar 
diffusion in the neutron star core (TD96;  Heyl \& Kulkarni 1998).  The 
rate of surface heating would be lower by a factor $\sim \gamma_e m_e/m_p$
if the current were carried by relativistic $e^\pm$ pairs of Lorentz 
factor $\gamma_e$.

A significant rate of surface heating leads to the interesting possibility
of an emission line at the surface ion cyclotron frequency.  
The magnetospheric charges are stopped mainly by ion collisions in
magnetic field much stronger than $B_{\rm QED}$.  The stopping depth
is higher than in the case of a non-magnetic atmosphere, and the
broad-band spectrum will, as a result, be
closer to a black body than in the weak-field regime analyzed by
Zel'dovich \& Shakura (1969) and Deufel, Dullemond, \& Spruit (2001). 

3. A significant optical depth to  resonant cyclotron
 scattering, $\tau_{\rm res}
\sim (v/c)^{-1}\,(B_\phi/B_\theta)_{\theta=\pi/2}$, is generated by the
current carriers at the magnetic equator.  This optical 
depth is anisotropic and approaches 
zero at the magnetic axis, where the equilibrium current density is 
$J(\theta)\propto\theta^2$.  When the twist is large,
$(B_\phi/B_\theta)_{\theta=\pi/2} = O(1)$,
photons will experience multiple resonant scattering, forming a high
energy non-thermal tail.  At a fixed $L_X$, the hardness of this tail 
will increase with the strength of the overall twist imparted to the 
magnetosphere, and so one obtains an explanation for the trend of
increasing spectral hardness with overall burst activity in the combined
population of SGR and AXP sources.   

In our self-similar model,
$\tau_{\rm res}$ is independent of the charge and mass of the particles,
as well as the resonant radius.  The large difference in the 
charge/mass ratio of electrons and ions therefore allows a magnetar to have
more than one scattering corona, localized at quite different radii:
$R \sim 50-100$ km for electrons and $\sim 10-20$ km for ions. 
Ions will resonantly scatter X-rays only below the (Doppler-shifted)
surface cyclotron frequency, and so the spectral tail generated by ion 
cyclotron scattering will be cut off at a much lower frequency than in the case
of resonant electron scattering.  Measurements of the persistent emission
above $\sim 30$ keV can test the relative importance of the two mechanisms.

The overall similarity in the luminosities of the thermal and 
non-thermal components of SGR and AXP spectra has a simple interpretation 
in this model, but would be more difficult to understand if the 
non-thermal emission arose from an independent radiative process 
(such as synchrotron or curvature emission) in the magnetosphere.  
For these sources, the mechanism of multiple cyclotron 
scattering also has significant advantages over non-resonant 
Comptonization in a thin surface layer which is heated by magnetospheric 
charges (Zel'dovich \& Shakura 1969;  Deufel et al. 2001).  Aside from 
the relatively soft spectrum of the surface emission expected in a 
strong magnetic field, it will also be noted that the intrinsically 
{\it brightest} soft gamma repeater 0526$-$66 has a 
{\it softer} spectrum than the other three SGRs
(each of which are $\sim 10-30$ times less luminous; Kulkarni et al. 2001).

4.  The observed pulse profile is strongly modified by resonant
cyclotron scattering.  Three effects enter here:
the strong anisotropy in the optical depth to electron cyclotron scattering,
the aspherical shape of the resonant surface, and the doppler beaming
of the scattered radiation resulting from the bulk motion of the
charge carriers.   In addition, the flux of thermal seed photons generated 
by the surface impact of magnetospheric charges is strongly anisotropic 
even in this self-similar model.  Not only is the surface current
inhomogeneous, but a larger fraction of the radiative flux will be carried 
by the O-mode (which is beamed along the local magnetic field) than
is the case in passively cooling neutron stars.  This second effect is
enhanced if the magnetospheric current is concentrated on extended 
field lines (e.g. TD96). 

5. Soft Gamma Repeater flares provide {\it prima facie} evidence for
sudden variations in the magnetic field, and therefore in the electrical
currents flowing outside the star (Thompson et al. 2000).
We have shown that the energy available
in the external field can maintain these currents for
as long as $\sim 30\,(B_{\rm pole}/10^{14}~{\rm G})$ yr, if the current
is supported by electrons and ions stripped from
the neutron star surface.  It is interesting to note, in this regard,
that SGR 0526$-$66, which has been quiescent as a burst source since 1983,
has a persistent X-ray spectrum which is strongly non-thermal but
at the same time significantly softer than the actively bursting
SGRs (Kulkarni et al. 2001).   This model also provides an
explanation for the simplified pulse profile observed in the
persistent emission of SGR 1900$+$14 following the 27 August 1998
giant flare, which was maintained even after the X-ray flux
returned to the baseline value observed before the flare.  We
obtain a valuable constraint on {\it how} the magnetic
field was modified during the flare:  the current flowing along
extended magnetic field lines actually increased during the flare,
suggesting that it was triggered by the release of sub-surface stresses.

6. Given a fixed polar magnetic field $B_{\rm pole}$, the observed
rate of spindown will grow as the external field is twisted
up, and there is an increase in the fraction of the field lines
which open out across the speed-of-light cylinder.  Equivalently,
the real surface polar field
is reduced by a factor 3 for a net twist $\Delta\phi_{\rm N-S} =1$
radian, in comparison with the dipole formula (Fig. \ref{bsurface}).

7. Consideration of the stability of a twisted, force-free magnetosphere
leads to the requirement that, close to the magnetic axis, the
current flows in the same direction on both
closed and open field lines.  Such an alignment of the currents
can be achieved only if the charge flow is pair-loaded on open field
lines -- which allows a differential drift between positive and negative
charges to maintain a current opposite to $\rho_{GJ}c\hat R$ in one
hemisphere.   There is, in turn, a limiting spin period beyond which
the rate of spindown can no longer be accelerated with respect to an
orthogonal vacuum dipole.  If the pair cascade cuts off at a spin period
comparable to the maximum observed in the known radio pulsar population,
then one obtains an explanation for the observed narrow distribution
($P = 6-12$ s) of AXP spins, and the similarly narrow distribution
($P = 5-8$ s) of SGR spins.  Direct evidence for such a decay in the
torque is provided by the anomalous pulsar 1E 2259$+$586, which has a
characteristic age at least a factor of 10 larger than the age of
the SNR in which it resides (Thompson et al. 2000).

\subsection{Relation between the SGRs, AXPs, and the Soft X-ray Pulsars}
We collect, in this section, the various threads which link our
model of non-potential neutron star magnetospheres to the
observed behavior of the Soft Gamma Repeaters, Anomalous X-ray
Pulsars, and their possible cousins, a growing group of Soft
X-ray Pulsars.

\subsubsection{Polar Dipole Fields of the SGRs/AXPs and their
Relation to Radio Pulsar Fields}

The polar magnetic fields\footnote{This polar field exceeds
by a factor 2 the average surface field usually quoted in the radio 
pulsar literature.} of the two rapidly spinning down
SGRs 1806$-$20 and 1900$+$14 (Kouveliotou et al. 1998, 1999, Woods et al.
2001b) are inferred to be $B_{\rm pole} = 1-3\times 10^{15}$ G
from the standard magnetic dipole formula.  These fields lie 
a factor of 10-30 above the strongest fields measured in the
radio pulsar population, $B_{\rm pole} = 10^{14}~{\rm G}$
(Camilo et al. 2000).  The polar fields inferred analogously for the AXPs
are continuous with
the pulsar distribution:  the source 1E 2259$+$586 has a long
characteristic age $P/2\dot P = 2.3\times 10^5$ yr and polar dipole field
$1.2\times 10^{14}$ G.  However, most of the AXPs have nominal dipole
fields 5-10 times this value.  

In our model, the actual polar magnetic fields of these sources will
lie below the classical magnetic dipole value.  A reduction of $\sim 3-10$ is
plausible in some sources (corresponding to net twist angles of $\sim 1-1.5$ radians).  The higher spindown rates
measured in the SGRs, as compared with the AXPs, could simply represent
a greater degree of magnetospheric twist imparted by deformations of
the magnetic field that are associated with bursting activity.  Indeed, the two AXPs with the fastest spindown also have the hardest persistent
X-ray spectra (e.g. Table 2 of Kaspi et al. 2001).  This positive correlation
between spindown rate and spectra becomes even stronger when the SGRs 
and AXPs are lumped together (Marsden \& White 2001).
Nonetheless, some distribution of surface
fields is almost certainly present in the AXP and SGR populations,
and it is plausible that some of the quiescent sources really do have 
weaker surface fields than the bursting sources.  

Additional physical constraints on the dipole fields of the SGR
sources come from these independent lines of argument:

i) Confinement of the relativistically hot plasma which powered
the pulsating tails of the two giant flares requires magnetic
fields stronger than $10^{14}\,(E/10^{44}~{\rm erg})^{1/2}$ G
(Thompson \& Duncan 1995).  This argument has been generalized
to allow for the possibility that the dipole is offset from
the center of the star, which would reduce the magnetic moment
corresponding to a fixed plasma energy $E$ (Thompson \& Duncan 2001).  
Even including this effect, one deduces $B_{\rm pole} \ga 10^{14}$ G, 
because a compact fireball would have a very high internal temperature
$T \gg 1$ MeV and would lose its energy rapidly to neutrino radiation
through $e^+ + e^- \rightarrow \nu + \bar\nu$ (instead of the observed
X-ray flux).

ii) Transient spindown of SGR 1900+14, $\Delta P/P = 1\times 10^{-4}$,
was observed within 3 months of the 27 August 1998 giant flare (Woods 
et al. 1999c).  A plausible mechanism for this torque involves a particle 
wind during the giant flare itself, combined with scattering of the 
X-rays by matter suspended 
in the magnetosphere near $R \sim 200$ km (where the momentum flux 
is high enough to break open the magnetic field lines).   If the
magnetic field were dipolar inside this `Alfv\'en' radius, then the maximum
torque would be $\Delta P/P \simeq 10^{-5}\,(B_{\rm pole}/10\,B_{QED})$
(Thompson et al. 2000).  However, if the magnetic field were twisted
close to the star, then the Alfv\'en radius would increase.  The net 
torque would be brought close to the observed value 
if $B_{\rm pole} \ga 5\times 10^{14}$ G (eq. [\ref{deltapp}]).

iii) The surface X-ray flux predicted by our self-similar model
(eq. [\ref{lsurf}]) is comparable to the observed luminosities
of the SGR and AXP sources if $B_{\rm pole} \sim 10^{14}$ G, but
is excessively large if $B_{\rm pole}$ is as large as $10^{15}$ G.
However, it should be emphasized that the internal stresses acting
on the crust of a magnetar may be more localized, and its entire 
magnetosphere need not be twisted.  The polar field could be as strong 
as $\sim 10^{15}$ G if the current were intermittent and flowed only 
over a fraction of the neutron star surface.

All of these arguments point to dipole fields stronger than $10^{14}$ G.  
A polar field several times the strongest pulsar field appears needed
to explain a transient spindown $\Delta P/P = 1\times 10^{-4}$ of SGR 
1900+14 by an outflowing wind during the 27 August flare.
The quiescent SGR 0526$-$66 has a softer
spectrum (Kulkarni et al. 2001).  We predict that its spindown rate,
when measured, will be intermediate between these sources and
the AXPs.

We re-emphasize that the actual surface fields of the SGR sources
are probably much stronger than the (corrected) dipole fields:
the complicated pulse profile observed during the 27 August flare
provides direct evidence for the presence of higher multipoles
(Feroci et al. 2001; Thompson \& Duncan 2001).  In addition, only magnetic
fields stronger than $\sim 4\times 10^{15}$ G will experience
rapid ambipolar diffusion through the core of a magnetar (TD96; Heyl 
\& Kulkarni 1998).

To summarize:  our model indicates that the distribution of true polar
(dipole) magnetic fields of the SGRs/AXPs is significantly narrower than
the classical dipole formula would suggest.  The range of spindown
rates is broadened by magnetospheric currents.  The true polar fields
of the SGRs and AXPs are continuous with the distribution of radio
pulsar fields; but are probably stronger than $10^{14}$ G in some sources.
Radio pulsars correspond to those sources in
which the toroidal field is either absent, or too weak to shear the
crust.  Furthermore
if an accelerated torque in the SGR/AXP sources is associated
with an active pair cascade on open field lines, then strong-B fields
do not, in themselves, suppress pair creation:   there must be a
greater similarity in this regard between active magnetars and
ordinary radio pulsars, than some calculations have suggested.

\subsubsection{Relation between the SGRs and AXPs}

The SGRs are distinguished from the AXPs by the emission of bright X-ray
outbursts and, in their quiescent states, harder X-ray spectra
and faster spindown.  We have shown that the last two properties
can be explained by a greater degree of twist in the external magnetic
field.  In addition, the sudden untwisting of an (internal) magnetic
field provides an attractive mechanism for powering the giant flares of
the SGRs.  This model suggests that some AXPs may be dormant SGRs:  
indeed, the SGRs go through long periods of quiescence
(the LMC source SGR 0526$-$66 has not been observed to burst since 1983;
Golenetskii et al. 1987).

Even in such a unified description of the SGRs and AXPs, the
question remains as to whether one type of activity typically
precedes the other in a given source, or whether instead an AXP will
undergo sporadic intervals of SGR activity which are separated by periods
of silence as a burst source.  In fact, some AXPs may never manifest
SGR behavior.
It has been suggested by Gaensler et al. (2001) that the AXPs are
characteristically younger than
the SGRs, because 3 of the 6 AXP sources are situated very close to the
centers of supernova remnants.  On the other hand, the other 3 AXPs do
not have obvious SNR counterparts, and one soft gamma repeater (SGR 1806$-$20) sits close to the center of the radio nebula G10.0$-$0.3 (Kulkarni et al. 1994).
Circumstantial evidence for high proper motions in two SGRs comes from
the projected position of SGR 0526$-$66 near the edge of the LMC remnant 
N49 (Cline 1982), and the position of SGR 1900$+$14 just outside SNR G42.8$+$0.6 (Hurley et al. 1999b).  A systematic difference
in proper motions between the AXPs and SGRs, if real, is an important clue to the conditions which give rise to these sources, but our model does not offer any unambiguous suggestion for what that difference may be.

The short spindown ages of SGRs 1900$+$14 and 1806$-$20 ($P/\dot P < 3000$
yrs: Kouveliotou et al. 1998, 1999; Woods et al. 2001c) seem to provide 
evidence, at first sight, that these sources are younger than most of 
the AXPs.  This impression could, however, be an artifact of a sufficiently 
strong twisting of the external field.  If the sources are in fact older
(as their positions with respect to the nearest SNR would suggest) then their
spindown must be persistently but {\it transiently} accelerated with respect
to the AXP population (Thompson et al. 2000).  Combining the two populations,
any AXP must spend at most $\sim 25$\% of
its $\sim 10^4-10^5$ yr lifetime as a bright X-ray source in an SGR
mode (Thompson et al. 2000).
However, the length of any given interval of SGR activity is poorly
constrained at present:  it must be at least $10-20$ yrs,
but could easily be much longer than that.  

We conclude that no unambiguous sequence of SGR and AXP
activity is discernable from the data, in part because the spindown
torques of the bursting sources appear to be accelerated transiently, and also because a wide range of proper motions
may exist in the combined population of SGRs and AXPs.
A plausible scenario is one in which
a portion of the AXP population undergoes intermittent periods of bursting
activity. It should be kept in mind that SGR activity may be concentrated during a particular range of ages, when the crust
of the star is colder and more brittle.  Because the spindown torque 
may be increased substantially by global magnetospheric currents,
the true polar fields of the actively bursting SGRs need not lie at the
extreme high end of the magnetar population.

This model is testable by long-term monitoring of the spin
of an SGR source, after it ends a period of bursting activity.  Thus,
continuous (phase-connected) monitoring of these sources is crucial to
the unraveling of the relation between the SGR and AXP phenomena.

\subsection{Connection between the SGRs and AXPs and nearby Soft X-ray Pulsars}

The nearby soft X-ray pulsars RX J0420.0$-$5022, RX J0720.4$-$3125,
and RBS 1223 have spin periods ($P = 22.7$, 8.37 and 5.2 s; Neuh\"auser
\& Tr\"umper 1999) remarkably close to the SGRs and AXPs.
It has been noted (Heyl \& Hernquist 1998; Kulkarni \& van Kerkwijk 1998) 
that these sources may be aged magnetars.  They are much fainter X-ray
sources than the SGRs and AXPs and could, at earlier times, have
been observable either as radio pulsars or as SGRs/AXPs.  
(The microphysical heating mechanism most plausibly is
a combination of ohmic decay and Hall deformations in the neutron
star crust, because ambipolar diffusion of a magnetic field through the core 
should be largely frozen when the surface temperature is as low as 
$\sim 60-100$ eV.)  If these objects are evolved from the SGR/AXP
population, then they provide further evidence for a decay of the torque 
beyond a characteristic spin period.    The recent possible detection of 
rapid spindown in the source RBS 1223 implies a surprisingly short 
characteristic age of $\sim 10^4$ yrs (Hambaryan et al. 2001), which 
is not consistent with this simple scenario.

\subsection{Transient Effects}

This model also provides a basis for interpreting time-dependent effects
observed in the SGR and AXP sources, including variations in flux and
pulse profile (Iwasawa, Koyama, \& Halpern 1992; Woods et al. 2001a), and
variations in torque (Paul et al. 2000; Kaspi et al. 2001; Woods et al.
2001b).  Indeed, these transient effects probably provide the strongest
constraints on the physical processes operating in the magnetospheres
of the SGR and AXPs.  

We are optimistic that progress in understanding
the electrodynamics of neutron stars with actively decaying magnetic
fields -- magnetars -- will occur more rapidly than has been the case with
radio pulsars -- for the simple reason that the observations place many
more direct constraints on theoretical models.

\acknowledgments
We thank Phil Arras, Vicky Kaspi, Chryssa Kouveliotou, and Peter Woods 
for comments and stimulating discussions.  CT acknowledges the support 
of the NSERC of Canada, and the Alfred P. Sloan Foundation. ML acknowledges
the support
of a CITA National Fellowship.  CT and ML also thank the Institute for 
Theoretical Physics at the University of California at Santa Barbara 
(NSF grant PHY99-0749) for its support during the workshop on
`Spin, Magnetism and Rotation in Young Neutron Stars', when part of
this work was done.  

\appendix
\section{Polarization Mode Exchange.}\label{polarization}
In this appendix we show that
even near the center of the cyclotron resonance, the dielectric 
properties of the magnetosphere are dominated by vacuum polarization.
The plasma contribution to the refractive index is
\begin{equation}\label{index}
|n-1|_{\rm plasma} = {2\pi Zen_Z c\over B\Delta\omega}
\end{equation}
at a frequency $\omega = ZeB/M c \pm \Delta\omega$.  (Here $Ze$, $M$ and
$n_Z$ are the charge, mass and density of the resonant particles, which 
could either be electrons or ions.)  Estimating $Ze n_Z \sim B/4\pi R$ 
through Amp\`ere's equation, this gives
\begin{equation}\label{indexb}
|n-1|_{\rm plasma} \sim {1\over 4\pi}\,\left({\lambda\over R}\right)
\left({\Delta\omega\over\omega}\right)^{-1},
\end{equation}
independent of $Z$ and $M$, but depending on the wavelength 
$\lambda = 2\pi c/\omega$ of the X-ray photon.  By contrast, 
the vacuum contribution to the index of refraction is
\begin{equation}\label{indexc}
|n-1|_{\rm vacuum} = K\alpha_{\rm em}^2\,\left({B\over B_{\rm QED}}\right)^2
\sin^2\theta = K\alpha_{\rm em}^2\,\left({\hbar\omega\over m_ec^2}\right)^2\,
\sin^2\theta,
\end{equation}
which we have evaluated at the electron cyclotron resonance 
$\hbar\omega = (B/B_{\rm QED})m_ec^2$.  In this expression 
$\alpha_{\rm em} = {1\over 137}$, and the constant $K = {7\over 90}$ 
for the O-mode and ${2\over 45}$ for the E-mode. 
The characteristic width of the resonance is determined by the thermal motion,
$\Delta\omega/\omega \sim (kT/m_ec^2)^{1/2} = 0.04
(kT/{\rm keV})^{1/2}$.  We conclude that the vacuum contribution 
is the larger by a factor $\sim 10^4\,
(\hbar\omega/{\rm keV})^3\,(\Delta\omega/\omega)$ at the electron 
cyclotron resonance (radius $R \sim 100$ km for magnetar-strength fields).  

The good photon polarization states are linear in this regime, 
and both are absorbed and emitted at the cyclotron resonance 
(which interacts with an elliptically polarized mode).  
The net result is that cyclotron scattering, at both ion and
electron resonances, will reduce the polarization of X-rays escaping
the star in a direction almost parallel to the local magnetic field,
and will induce a net linear polarization of X-rays escaping 
across closed magnetic field lines (with the polarization vector
lying perpendicular to {\bf B}).  Thus, measurements of the emergent
X-ray polarization (Heyl \& Shaviv 2000) will provide a direct probe of
the current flowing through the atmosphere of a magnetar.

\section{Resonant Cyclotron Force Vs. Gravity}\label{compforce}

The radiation field is anisotropic everywhere in the magnetosphere,
and a particle at rest will feel a force from resonant cyclotron scattering.
We now show that if the surface X-ray flux is powered self-consistently by
the impact of magnetospheric charges (\S \ref{surfheat}), then the
radiative force acting on electrons is typically large compared with
gravity.  The conclusion is slightly more complicated for ions:  in the
quiescent state of an SGR or AXP source, the radiative force acting on
them is typically small compared with gravity at the {\it surface} of
the star.  The radiative force remains weaker than gravity at greater
distances if the spectral intensity $L_\omega$ increases with frequency
below $\hbar\omega \sim 1$ keV ($L_\omega \sim \omega^\alpha$ with
$\alpha > 0$).

\subsection{Radiative Force}

The radiative force is easily estimated when the magnetic field is purely
radial, and the radiation field is axially symmetric about {\bf B}.
Because the re-emitted photon carries vanishing average momentum,
this force is
\begin{equation}\label{frad}
F_{\rm rad} = \int d\omega \int_0^1 2\pi d(\cos\theta)
\left[{dL_\omega\over d\Omega}\cos\theta\right]\,
{\sigma_{\rm res}(\omega,\cos\theta)\over 4\pi R^2c}
\end{equation}
(we neglect the effect of the recoil;  cf. Sincell \& Krolik 1992).
Substituting $\sigma_{\rm res}(\omega,\cos\theta) =
(Z^2\pi^2e^2/Mc)\,(1+\cos^2\theta)\,\delta(\omega-\omega_c)$ 
near the cyclotron frequency $\omega_c = ZeB/Mc$, one finds 
(see Mitrofanov \& Pavlov 1982 for the case of electron scattering)
\begin{equation}\label{fradb}
F_{\rm rad} = K_{\rm rad}{Z\pi^2e^2\over eB}
{(\omega L_\omega)_{\omega_c}\over 4\pi R^2c}.
\end{equation}
(The numerical coefficient $K_{\rm rad} = 2$ in the case of a purely radial
photon field; whereas $K_{\rm rad} = {3\over 4}$ near the surface of a
black body.)  Comparing with the gravitational force $F_{\rm grav} =
GMM_{\rm NS}/R^2$ on the scattering charge, and substituting eq.
(\ref{lsurf}) for $L_X = \int L_\omega d\omega$, one finds
\begin{equation}\label{fradc}
{F_{\rm rad}\over F_{\rm grav}} = {\pi K_FZ\,p\,F(\pi/2)\over 8(2+p)}\,
\left({M\over m_p}\right)^{-1}\,
\left({B\over B_{\rm pole}}\right)^{-1}\,
\left[{(\omega L_\omega)_{\omega_c}\over L_X}\right]
\left({B_\phi\over B_\theta}\right)_{\theta = \pi/2}.
\end{equation}
The numerical coefficient in front is $O(10^{-1})$, and for electrons
one deduces
\begin{equation}\label{fradd}
{F_{\rm rad}\over F_{\rm grav}} \sim 10^2\,
\left({R_{\rm res}\over R_{\rm NS}}\right)^3\,
\left[{(\omega L_\omega)_{\omega_c}\over L_X}\right]\,
\left({B_\phi\over B_\theta}\right)_{\theta = \pi/2} \gg 1
\;\;\;\;\;\;({\rm electrons}).
\end{equation}
(If the twist is very small, then internal heating can easily power
a large enough surface X-ray flux to enforce the same inequality.)

By contrast, ions will interact resonantly with X-rays at the surface
of the star in the presence of $\sim 10^{14}-10^{15}$ G magnetic fields,
and one deduces
\begin{equation}\label{frade}
{F_{\rm rad}\over F_{\rm grav}} \sim 0.1\,\left({Z\over A}\right)\,
\left[{(\omega L_\omega)_{\omega_c}\over L_X}\right]\,
\left({B_\phi\over B_\theta}\right)_{\theta = \pi/2} \la 1
\;\;\;\;\;\;(\rm ions).
\end{equation}
(Atoms of large atomic number $Z$ will become bound in molecular chains in 
magnetar-strength fields:  Lai \& Salpeter 1997;  Thompson et al. 2000.)

The radiation force on the ions can overcome gravity at the surface of
the star only if the luminosity $(\omega L_\omega)_{\omega_{c,p}}$
of X-rays at the surface cyclotron frequency is much higher than 
eq. (\ref{lsurf}).  One possible example of such a situation is the 
burst from SGR 1900$+$14 on 29 August 1998, which showed an extended faint 
tail of X-ray emission with a very hard spectrum  ($kT \ga 4$ keV).
That tail may represent transient cooling of a relatively small
surface hotspot following the dissipation of a very hot magnetospheric
plasma (Ibrahim et al. 2001).  The radiative force at the
line would also be increased in the presence of
surface heating by magnetospheric charges;  but the mildly relativistic
upward motion of the ions would quickly shift them out of resonance with 
the surface cyclotron frequency.
Finally, it has been suggested that in
some magnetar candidates, the steep high energy spectral tail extends
to frequencies well below $\hbar\omega \sim 1$ keV (Kulkarni et al. 2001).
In that case, it would be possible for the radiative force on the ions
to exceed gravity beyond a certain distance from the star.

\subsection{Draining Suspended Material through Persistent Currents}

Notice that in the case of a neutral ion-electron plasma, the radiative
force on the electrons indicated by eq. (\ref{fradd}) could exceed
gravity by more than a factor $\sim (A/Z)(m_p/m_e)$, thereby allowing
plasma which is blown into the magnetosphere during an X-ray flare
to be supported in the magnetosphere against gravity as the X-ray flux
returns to the baseline value.  However, the mass of plasma which can be
so supported (in e.g. a thin disk near the magnetic equator: 
Zheleznyakov \& Serber 1994) is small enough that it would 
quickly be drained by even a relatively weak magnetospheric current. 
Balancing the radiation pressure normal to the `disk' against the
normal component of gravity at a scale height $h$ above the magnetic equator,
\begin{equation}\label{forcebal}
{(\omega L_\omega)_{\omega_c}\over 4\pi R^2 c}\,\left({h\over R}\right)
\la \Sigma{GM_{\rm NS}\over R^2}\,\left({h\over R}\right),
\end{equation}
gives the maximum surface density which can be supported against gravity
by resonant scattering,
\begin{equation}\label{sigmax}
\Sigma \la {(\omega L_\omega)_{\omega_c}\over 4\pi GM_{\rm NS} c}.
\end{equation}
(This estimate requires a large optical depth to resonant scattering
$\tau_{\rm res} \sim F_{\rm rad}/F_{\rm grav}$, which can indeed be 
maintained when the radiative force $F_{\rm rad}$ acting on the electrons
is stronger than gravity.)

The flux of charges across the magnetic equator at radius $R$ is
$J/e \sim (c/4\pi)(2B_\theta/R)\,(B_\phi/B_\theta)_{\theta=\pi/2}$.  If a
source of charges is available in the magnetosphere, it is
energetically favorable for the current to tap it (instead of being
drawn from charges lifted off the neutron star surface).  Thus, one
can expect the `disk' to be drained by the current, in a very short
time:
\ba
t_{\rm drain} = {2\Sigma/m_p\over J/e} &\sim&
{R_{\rm NS}\over c}\,\left({(\omega L_\omega)_{\omega_c}\over
GM_{\rm NS}B_{\rm pole}m_p c/e}\right)\,\left({R\over R_{\rm NS}}\right)^4\,
\left({B_\phi\over B_\theta}\right)_{\theta = \pi/2}^{-1}
\nn
&\sim& 0.03\,\left[{(\omega L_\omega)_{\omega_c}\over L_X}\right]\,
\left({R\over 10\,R_{\rm NS}}\right)^4\;\;\;\;\;{\rm s}.
\label{tdrain}
\ea
Notice that this expression has no explicit dependence on the magnitude
of the twist, if the surface X-ray flux is powered self-consistently
by the impact of magnetospheric charges. 
	
Matter can also be suspended by the centrifugal force beyond the corotation
radius,
\begin{equation}\label{rco}
 R_{\rm co} = (GM_{\rm NS})^{1/3}\,\Omega_{\rm NS}^{-2/3},
\end{equation}
 even if the radiative force is weaker than gravity.  The maximum
column density which can be so suspended can be estimated by balancing the
ram pressure of material corotating with the star, with the local
 magnetic tension (Thompson 2000; Ibrahim et al. 2001),
\begin{equation}\label{ebal}
{\Sigma\over 2h}(R_{\rm co}\Omega_{\rm NS})^2 \simeq
{B_\theta^2(R_{\rm co})\over 8\pi}.
\end{equation}
Here $h$ is the scale height of the plasma above the magnetic equator.
Substituting eq. (\ref{rco}), and re-expressing $\Sigma$ in terms of
a Thomson optical depth, we obtain
\begin{equation}\label{disktau}
\tau_T \simeq {\Sigma\sigma_T\over m_p} = 2\,
\left({B_{\rm pole}\over 10^{15}~{\rm G}}\right)^2\,
\left({h/R_{\rm co}\over 10^{-2}}\right)\,
\left({P\over 6~{\rm s}}\right)^{-8/3}
\end{equation}
(for $M_{\rm NS} = 1.4\,M_\odot$ and $R_{\rm NS} = 10$ km).  The
scale height can be estimated as
\begin{equation}\label{hrratio}
{h(R_{\rm co})\over R_{\rm co}}
\sim \left({kT\over m_pg(R_{\rm co})R_{\rm co}}\right)^{1/2}
\end{equation}
where the temperature of the plasma is reduced from the surface
value by the factor $T/T_S \sim (h/R_{\rm co})^{1/4}\,
(R_{\rm co}/R_{\rm NS})^{-1/2}$.  This gives
\begin{equation}\label{scaleheight}
{h(R_{\rm co})\over R_{\rm co}} \sim
0.004\,\left({kT_S\over 0.5~{\rm keV}}\right)^{4/7}\,
\left({P\over 6~{\rm s}}\right)^{4/21}.
\end{equation}

The normalization $B_{\rm pole} \sim 10^{15}$ G of the surface dipole field
in eq. (\ref{disktau}) is appropriate to the actively bursting SGRs 1806$-$20 and
1900$+$14, if they have purely dipolar magnetic fields (Kouveliotou et al. 1998,
1999).  However, we have seen (in \S \ref{spindown}) that the real
value of $B_{\rm pole}$ is reduced if the magnetic field decreases with 
radius more slowly than a dipole, ${\bf B} \sim R^{-(2+p)}$ with $p < 1$.  This
also has the effect of increasing the mass which can be contained by 
the magnetic tension against the centrifugal force.  For a fixed
$P$ and $\dot P$,  the net effect is to
re-scale $B_{\rm pole}$ from the magnetic dipole value by the factor
\begin{equation}\label{bscale}
B_{\rm pole} \rightarrow \left({\Omega R_{\rm co}\over c}\right)^{1-p}
\;B_{\rm pole}({\rm MDR}) < B_{\rm pole}({\rm MDR})
\end{equation}
This factor is $0.55\,(P/6~{\rm s})^{-0.05}$ in the case $p = 0.85$.
Combining it with eq. (\ref{scaleheight}) in eq. (\ref{disktau}), we
conclude that the suspended matter can maintain a modest optical depth
to Thomson scattering, $\tau_T > 1$, when $B_{\rm pole}({\rm MDR}) \ga
2\times 10^{15}$ G.
  
Finally, let us write down the time for a persistent current to drain this
suspended material.  It is
\begin{equation}\label{tdrainb}
t_{\rm drain} = 4\,\left({B_{\rm pole}\over 10^{15}~{\rm G}}\right)\,
\left({h/R_{\rm co}\over 10^{-2}}\right)\,
\left({B_\phi\over B_\theta}\right)_{\theta = \pi/2}^{-1}\;\;\;\;\;\;{\rm yr},
\end{equation}
assuming $M_{\rm NS} = 1.4\,M_\odot$ and $R_{\rm NS} = 10$ km.
Substituting once again eqs. (\ref{scaleheight}) and (\ref{bscale}),
one infers a drainage time $t_{\rm drain} \sim 1\,
(B_{\rm pole}/10^{15}~{\rm G})\,(B_\phi/B_\theta)_{\theta = \pi/2}^{-1}$ yr
(quoted here for $p = 0.85$).  Of course, the density of suspended
material may be high enough for it to spin outward from the star
through the action of the centrifugal force, and settle into
a rotationally supported disk (Thompson 2000).

\clearpage
\figurenum{1}

\clearpage
\figurenum{1}
\begin{figure}
\caption{A twisted magnetic field is anchored in the highly conducting
interior of a degenerate star.   The twist is initially
confined to the interior of the star, so that the current closes
at the surface by flowing across the magnetic field.  The resulting
${1\over c}{\bf J}\times{\bf B}$ force causes the liquid near the
surface to rotate, so as to distribute the twist more uniformly
along the magnetic field lines.  The net effect is to force the
current to flow out of the star, into its `magnetosphere'.
In the case of a magnetar, this process may be partly stabilized
by the rigidity of the crust, so that the external field twists
up intermittently (giving rise to SGR flares).
}
\end{figure}

\clearpage

\clearpage

\clearpage

\clearpage

\clearpage
\figurenum{5}
\begin{figure}
\caption{In one magnetic hemisphere, a space-charge limited flow
will generate a current of the opposite sign to the current flowing
on the closed, twisted field lines.  The toroidal component of
the field will then have an opposing sense on the open and
closed field lines.  In the absence of pairs, this may cause a 
cancellation of the twist in the intermediate regions of the magnetosphere.
An electron-positron pair cascade allows the density of current-carrying 
particles to build up, so that the Goldreich-Julian charge density 
$-{\bf\Omega}\cdot{\bf B}/2\pi c$ can be maintained even as the current 
flows {\it backward} compared with a pure space-charge limited flow.
}
\end{figure}

\clearpage

\appendix
\end{document}